\documentclass[prd,aps,floats,amssymb,preprint,nofootinbib]{revtex4}

\usepackage{epsfig}
\usepackage{amssymb}
\usepackage{bbm}
\usepackage{color}
\usepackage{ulem}

\definecolor{MyDarkBlue}{rgb}{0.15,0.15,0.45}
\usepackage[linktocpage=true]{hyperref}
\hypersetup{
colorlinks=true,
citecolor=MyDarkBlue,
linkcolor=MyDarkBlue,
urlcolor=MyDarkBlue,
pdfauthor={Garrett Goon, Kurt Hinterbichler and Mark Trodden},
pdftitle={General Embedded Brane Effective Field Theories},
pdfsubject={hep-th}
}

\usepackage{enumitem}
\usepackage{graphicx}
\usepackage{subfigure}
\usepackage{amsmath}
\usepackage{bbm}
\usepackage{color}

\newcommand{\be}{\begin{equation}}
\newcommand{\ee}{\end{equation}}
\newcommand{\bea}{\begin{eqnarray}}
\newcommand{\eea}{\end{eqnarray}}
\newcommand{\beas}{\begin{eqnarray*}}
\newcommand{\eeas}{\end{eqnarray*}}
\newcommand{\nn}{\nonumber}

\newcommand{\sech}{\rm sech}
\newcommand{\csch}{\rm csch}
\newcommand{\pir}{\left(\pi\over {\cal R}\right)}
\newcommand{\pirt}{\left(2\pi\over {\cal R}\right)}

\def\({\left(}
\def\){\right)}
\newcommand{\rc}{\mathcal{R}}
\newcommand{\Ac}{\mathcal{A}}
\newcommand{\Bc}{\mathcal{B}}

\newcommand{\partialb}{\bar{\partial}}
\newcommand{\half}{\frac{1}{2}}

\begin{document}

%=====TITLE AND ABSTRACT===============================================================

\title{Symmetries for Galileons and DBI scalars on curved space}
\author{Garrett Goon, Kurt Hinterbichler\footnote{kurthi@physics.upenn.edu} and Mark Trodden\footnote{trodden@physics.upenn.edu}}

\affiliation{Center for Particle Cosmology, Department of Physics and Astronomy, University of Pennsylvania,
Philadelphia, Pennsylvania 19104, USA}

\date{\today}

\begin{abstract}
We introduce a general class of four-dimensional effective field theories which include curved space Galileons and DBI theories possessing nonlinear shift-like symmetries. These effective theories arise from purely gravitational actions for 3-branes probing higher dimensional spaces. In the simplest case of a Minkowski brane embedded in a higher dimensional Minkowski background, the resulting four-dimensional effective field theory is the Galileon one, with its associated Galilean symmetry and second order equations.  However, much more general structures are possible.  We construct the general theory and explicitly derive the examples obtained from embedding maximally symmetric branes in maximally symmetric ambient spaces.  Among these are Galileons and DBI theories with second order equations that live on de Sitter or anti-de Sitter space, and yet retain the same number of symmetries as their flat space counterparts, symmetries which are highly non-trivial from the $4$d point of view.  These theories have a rich structure, containing potentials for the scalar fields, with masses protected by the symmetries. These models may prove relevant to the cosmology of both the early and late universe.
\end{abstract}

\maketitle

\tableofcontents

\setcounter{footnote}{0}

\section{\label{introduction}Introduction and Summary}
The possibility that the universe may contain large, and possibly infinite, spatial dimensions beyond the three we commonly perceive has opened up entirely new avenues to address fundamental questions posed by particle physics and by cosmology. The precise manner in which the dynamics of the higher-dimensional space manifests itself in the four dimensional world depends on the geometry and topology of the extra-dimensional manifold, and the matter content and action chosen. At low enough energies, the relevant physics is then captured by a four-dimensional effective field theory with properties inherited from the specific higher-dimensional model under consideration. The simplest example of this is the Kaluza-Klein tower -- the hierarchy of higher mass states that accompany zero mass particles when compactifying a five-dimensional theory on a circle. There are, however, much more exotic possibilities.  Many of these describe viable higher-dimensional theories, while others are merely mathematical tools with which to construct interesting physical four-dimensional effective field theories.

A particularly interesting and well studied example of a higher-dimensional model is the Dvali-Gabadadze-Poratti (DGP) model \cite{Dvali:2000hr}, for which the ambient space is a flat $5$-dimensional spacetime in which a Minkowski $3$-brane floats, subject to an action consisting merely of two separate Einstein Hilbert terms -- one in $5$d, and the other only on the brane, constructed from the induced metric there.  In an appropriate limit, the resulting four-dimensional effective field theory describes gravity plus a scalar degree of freedom parametrizing the bending of the brane in the extra dimension \cite{Luty:2003vm,Nicolis:2004qq}. The specific form of the four dimensional action for the scalar inherits a symmetry from a combination of five dimensional Poincar\'e invariance and brane reparametrization invariance. In the small field limit this symmetry takes a rather simple form and has been called the {\it Galilean} symmetry, with the associated scalar becoming the {\it Galileon} \cite{Nicolis:2008in}.  

Abstracting from DGP, a four dimensional field theory with this Galilean symmetry is interesting in its own right.  It turns out that there are a finite number of terms, the {\it Galileon terms}, that have fewer numbers of derivatives per field than the infinity of competing terms with the same symmetries.  These terms have the surprising property that, despite the presence of higher derivatives in the actions, the equations of motion are second order, so that no extra degrees of freedom are propagated around any background.  Much has been revealed about the Galileon terms, including such useful properties as a non-renormalization theorem \cite{Luty:2003vm,Hinterbichler:2010xn,Burrage:2010cu}, and applications in cosmology \cite{Agarwal:2011mg,Burrage:2010cu,Creminelli:2010ba,Creminelli:2010qf,DeFelice:2010as,Deffayet:2010qz,Kobayashi:2011pc,Mota:2010bs,Wyman:2011mp}.  The Galileons have been covariantized \cite{Deffayet:2009mn,Deffayet:2009wt,Deffayet:2011gz}, extended to p-forms \cite{Deffayet:2010zh}, and supersymmetrized \cite{Khoury:2011da}.  Further, it was recently shown that the general structure of Galileon field theories can be extended to multiple fields, finding their origins in braneworld constructions with more than one codimension \cite{Hinterbichler:2010xn,Padilla:2010de,Padilla:2010ir,Padilla:2010tj,Zhou:2010di}. If some of the resulting symmetries of the four dimensional effective field theory are broken, then they are related to low energy descriptions of cascading gravity models in which a sequence of higher dimensional branes are embedded within one another \cite{deRham:2007rw,deRham:2007xp,Agarwal:2011mg,Agarwal:2009gy}.

If our universe really is a brane world, then theories of this sort are generic, since they share, in a certain limit, the symmetries of the Dirac-Born-Infeld (DBI) action.  The DBI action encodes the lowest order dynamics of a brane embedded in higher dimensions, and provides an 
important arena within which to study inflation~\cite{Silverstein:2003hf,Alishahiha:2004eh}, late-time cosmic acceleration~\cite{Ahn:2009xd}, tunneling~\cite{Brown:2007zzh}, and exotic topological defects~\cite{Andrews:2010eh,Babichev:2006cy,Sarangi:2007mj,Bazeia:2007df,Babichev:2008qv}.   The Galileon terms can be thought of as a subset of the higher order terms expected to be present in any effective field theory of the brane, and which will be suppressed by powers of some cutoff scale.  The Galileons are a special subset in the class of all possible higher order terms because they contain fewer derivatives per field than competing terms with the same symmetries, and because they yield second order equations.  Crucially, there can exist regimes in which only a finite number of Galileon terms are important, and the infinity of other possible terms within the effective field theory are not (see section II of \cite{Hinterbichler:2010xn}, as well as \cite{Nicolis:2004qq,Endlich:2010zj}, for more on this and for examples of such regimes.)  This fact, coupled with a non-renormalization theorem for Galileons and the fact that there are a finite number of such terms, holds out the hope of computing non-linear facts about the world which are exact quantum mechanically.   Finally, it should be remembered that even if our universe is not a brane world, the same conclusions follow if one postulates the existence of symmetries of the same form as those of a brane world.

In this paper, we construct a general class of four-dimensional effective field theories by writing an action on a 3-brane probing a higher dimensional bulk, of which the Galileon theory and DBI scalars are special cases. This extends the construction of~\cite{deRham:2010eu} to its most general form.  We observe that the symmetries inherited by scalar fields in the $4$d theory are determined by isometries of the bulk metric, and are present if and only if the bulk has isometries.  The precise manner in which the symmetries are realized is determined by the choice of gauge, or foliation, against which brane fluctuations are measured.  We derive in general the symmetries of these effective field theories, and classify the examples that result when embedding a maximally symmetric brane in a maximally symmetric background.  This approach yields a set of new Galileon-like theories which live on $4$d curved space but retain the same number of non-linear shift-like symmetries as the flat-space Galileons or DBI theories.  

These theories have their own unique properties.  For example, in curved space the field acquires a potential which is fixed by the symmetries -- something that is not allowed for the flat space Galileons.  In particular, the scalars acquire a mass of order the inverse radius of the background, and the value of the mass is fixed by the nonlinear symmetries.  Although not addressed in detail here, allowing for de Sitter solutions on the brane opens up the possibility of adapting these new effective field theories to cosmological applications such as inflation or late time cosmic acceleration in such a way that their symmetries ensure technical naturalness.

The paper is structured as follows. In the next section we discuss general brane actions and symmetries, and the ways in which these symmetries may be inherited by a four-dimensional effective field theory. In section~\ref{sec:ghostfreeactions} we then consider constructing actions with second order equations and explicitly derive all possible terms in such theories. We then provide six separate examples, exhausting all the maximally symmetric possibilities: a $4$d Minkowski brane embedded in a Minkowski bulk; a $4$d Minkowski brane embedded in $AdS_5$; a $4$d de Sitter brane embedded in a Minkowski bulk; a $4$d de Sitter brane embedded in $dS_5$; a $4$d de Sitter brane embedded in $AdS_5$; and a $4$d Anti-de Sitter brane embedded in $AdS_5$. In each case, we describe the resulting $4$d effective field theories and comment on their structure. In section~\ref{sec:smallfieldlimits} we take the small field limits to obtain Galileon-like theories, discuss their stability, and compare and contrast these theories with the special case of the original Galileon, before concluding.

{\bf Conventions and notation}:

We use the mostly plus metric signature convention.  The 3-brane worldvolume coordinates are $x^\mu$, $\mu=0,1,2,3$, bulk coordinates are $X^A$, $A=0,1,2,3,5$.  Occasionally we use 6-dimensional cartesian coordinates $Y^\Ac$, $\Ac=0,1,2,3,4,5$, for constructing five dimensional $AdS_5$ and $dS_5$ as embeddings. Tensors are symmetrized and anti-symmetrized with unit weight, i.e $T_{(\mu\nu)}=\half \left(T_{\mu\nu}+T_{\nu\mu}\right)$,   $T_{[\mu\nu]}=\half \left(T_{\mu\nu}-T_{\nu\mu}\right)$.  

When writing actions for a scalar field $\pi$ in curved space with metric $g_{\mu\nu}$ and covariant derivative $\nabla_\mu$, we use the notation $\Pi$ for the matrix of second derivatives $\Pi_{\mu\nu}\equiv\nabla_{\mu}\nabla_\nu\pi$.  For traces of powers of $\Pi$ we write $[\Pi^n]\equiv Tr(\Pi^n)$, e.g. $[\Pi]=\nabla_\mu\nabla^\mu\pi$, $[\Pi^2]=\nabla_\mu\nabla_\nu\pi\nabla^\mu\nabla^\nu\pi$, where all indices are raised with respect to $g^{\mu\nu}$.  We also define the contractions of powers of $\Pi$ with $\nabla\pi$ using the notation $[\pi^n]\equiv \nabla\pi\cdot\Pi^{n-2}\cdot\nabla\pi$, e.g. $[\pi^2]=\nabla_\mu\pi\nabla^\mu\pi$, $[\pi^3]=\nabla_\mu\pi\nabla^\mu\nabla^\nu\pi\nabla_\nu\pi$, where again all indices are raised with respect to $g^{\mu\nu}$.

\section{General brane actions and symmetries}
We begin with a completely general case - the theory of a dynamical 3-brane moving in a fixed but arbitrary (4+1)-dimensional background.  
The dynamical variables are the brane embedding $X^A(x)$, five functions of the world-volume coordinates $x^\mu$.  

The bulk has a fixed background metric $G_{AB}(X)$.  From this and the $X^A$, we may construct the induced metric $\bar g_{\mu\nu}(x)$ and the extrinsic curvature $K_{\mu\nu}(x)$, via
\bea 
\bar g_{\mu\nu}&=&e^A_{\ \mu}e^B_{\ \nu} G_{AB}(X), \\ 
K_{\mu\nu}&=&e^A_{\ \mu}e^B_{\ \nu}\nabla_A n_B \ .
\eea
Here $e^A_{\ \mu}= {\partial X^A\over\partial x^\mu}$ are the tangent vectors to the brane, and $n^A$ is the normal vector, defined uniquely (up to a sign) by the properties that it is orthogonal to the tangent vectors $e^A_{\ \mu}n^BG_{AB}=0$, and normalized to unity $n^An^BG_{AB}=1$.  (Note that the extrinsic curvature can be written $K_{\mu\nu}=e^B_{\ \nu}\partial_\mu n_B-e^A_{\ \mu}e^B_{\ \nu}\Gamma^C_{AB}n_C$, demonstrating that it depends only on quantities defined directly on the brane and their tangential derivatives.)

We require the world-volume action to be gauge invariant under reparametrizations of the brane,
\be 
\label{gaugetransformations} 
\delta_g X^A=\xi^\mu\partial_\mu X^A \ ,
\ee
where $\xi^\mu(x)$ is the gauge parameter.  This requires that the action be written as a diffeomorphism scalar, $F$, of $\bar g_{\mu\nu}$ and $K_{\mu\nu}$ as well as the covariant derivative $\bar\nabla_\mu$ and curvature $\bar R^\alpha_{\ \beta\mu\nu}$ constructed from $\bar g_{\mu\nu}$,
\be
\label{generalaction} 
S= \int d^4x\ \sqrt{-\bar g}F\left(\bar g_{\mu\nu},\bar\nabla_\mu,\bar R^{\alpha}_{\ \beta\mu\nu},K_{\mu\nu}\right) \ .
\ee

This action will have global symmetries only if the bulk metric has Killing symmetries.  If the bulk metric has a Killing vector $K^A(X)$, i.e. a vector satisfying the Killing equation
\be \label{killingequation}
K^C\partial_C G_{AB}+\partial_AK^CG_{CB}+\partial_BK^CG_{AC}=0 \ ,
\ee
then the action will have the following global symmetry under which the $X^A$ shift,
\be
\label{generalsym} 
\delta_K X^A=K^A(X) \ .
\ee
 It is straightforward to see that the induced metric and extrinsic curvature, and hence the action~(\ref{generalaction}), are invariant under~(\ref{generalsym}).  

\begin{figure} %  figure placement: here, top, bottom, or page
   \centering
   \includegraphics[width=4.0in]{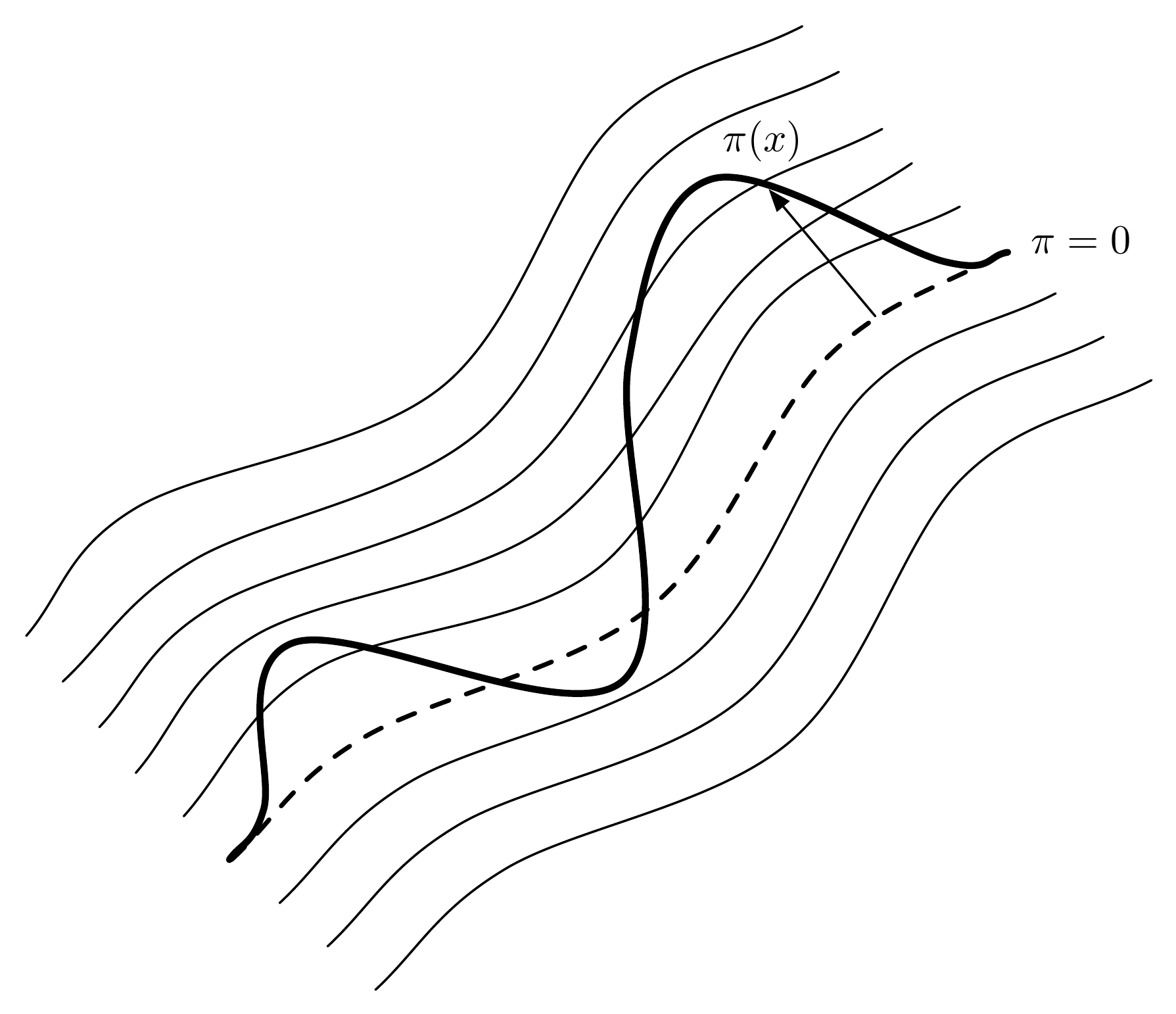}
   \caption{The field $\pi$ measures the brane position with respect to some chosen foliation.}
   \label{foliation}
\end{figure}

We are interested in creating non-gauge theories with global symmetries from the transverse fluctuations of the brane, so we now fix all the gauge symmetry of the action. We accomplish this by first choosing a foliation of the bulk by time-like slices.  We then choose bulk coordinates such that the foliation is given by the surfaces $X^5= {\rm constant}$.  The remaining coordinates $X^\mu$ can be chosen arbitrarily and parametrize the leaves of the foliation.  The gauge we choose is
\be
\label{physgauge} 
X^\mu(x)=x^\mu, \ \ \ X^5(x)\equiv \pi(x) \ .
\ee
In this gauge, the world-volume coordinates of the brane are fixed to the bulk coordinates of the foliation.  We call the remaining unfixed coordinate $\pi(x)$, which measures the transverse position of the brane relative to the foliation (see Figure~\ref{foliation}).  This completely fixes the gauge freedom.  The resulting gauge fixed action is then an action solely for $\pi$,
\be
\label{gaugefixedaction} 
S= \int d^4x\ \left. \sqrt{-\bar g}F\left(\bar g_{\mu\nu},\bar\nabla_\mu,\bar R^{\alpha}_{\ \beta\mu\nu},K_{\mu\nu}\right)\right|_{X^\mu=x^\mu,\ X^5=\pi} \ .
\ee

Global symmetries are physical symmetries that cannot be altered by the unphysical act of gauge fixing.  Thus, if the original action~(\ref{generalaction}) possesses a global symmetry (\ref{generalsym}), generated by a Killing vector $K^A$, then the gauge fixed action~(\ref{gaugefixedaction}) must also have this symmetry.   However, the form of the symmetry will be different because the gauge choice will not generally be preserved by the global symmetry.  The change induced by $K^A$ is
\be 
\delta_ Kx^\mu=K^\mu(x,\pi),\ \ \ \delta_K\pi=K^5(x,\pi) \ .
\ee
To re-fix the gauge to~(\ref{physgauge}), it is necessary to simultaneously perform a compensating gauge transformation with gauge parameter 
\be 
\xi_{\rm comp}^{\mu}=-K^\mu(x,\pi) \ .
\ee
The combined symmetry acting on $\pi$,
\be
\label{gaugefixsym} 
(\delta_K+\delta_{g,{\rm comp}})\pi=-K^\mu(x,\pi)\partial_\mu\pi+K^5(x,\pi) \ ,
\ee
is then a symmetry of the gauge fixed action~(\ref{gaugefixedaction}).  

\subsection{\label{maxsymsubsection} A special case}

We now specialize to a case which includes all the maximally symmetric examples of interest to us in this paper.  This is the case where the foliation is Gaussian normal with respect to the metric $G_{AB}$, and the extrinsic curvature on each of the leaves of the foliation is proportional to the induced metric.  With these restrictions, the metric takes the form
\be 
\label{metricform} 
G_{AB}dX^AdX^B=d\rho^2+f(\rho)^2g_{\mu\nu}(x)dx^\mu dx^\nu \ ,
\ee
where $X^5=\rho$ denotes the Gaussian normal transverse coordinate, and $g_{\mu\nu}(x)$ is an arbitrary brane metric.  Recall that in the physical gauge (\ref{physgauge}), the transverse coordinate of the brane is set equal to the scalar field, $\rho(x)=\pi(x)$.  

Working in the gauge (\ref{physgauge}), the induced metric is
\be 
\bar g_{\mu\nu}=f(\pi)^2g_{\mu\nu}+\nabla_\mu\pi\nabla_\nu\pi \ .
\ee
Defining the quantity
\be 
\gamma={1\over \sqrt{1+{1\over f^2}(\nabla\pi)^2}} \ ,
\ee
the square root of the determinant and the inverse metric may then be expressed as
\be 
\sqrt{-\bar g}=\sqrt{-g}f^4\sqrt{1+{1\over f^2}(\nabla\pi)^2}=\sqrt{-g}f^4{1\over \gamma},
\ee
and
\be 
\bar g^{\mu\nu}={1\over f^2}\left(g^{\mu\nu}-\gamma^2{\nabla^\mu\pi\nabla^\nu\pi\over f^2}\right) \ .
\ee
The tangent vectors are
\be 
e^A_{\ \mu}={\partial X^A\over \partial x^\mu}=\begin{cases}\delta^\nu_\mu & A=\nu \\ \nabla_\mu \pi & A=5\end{cases} \ .
\ee
To find the normal vector $n^A$ we solve the two equations
\bea 0&=&e^A_{\ \mu}n^BG_{AB}=f^2n^\nu g_{\mu\nu}+n^5\partial_\mu\pi, \\
1&=&n^An^BG_{AB}={1\over f^2}g^{\mu\nu}\partial_\mu\pi\partial_\nu\pi(n^5)^2+(n^5)^2 \ ,
\eea
to obtain
\be 
n^A=\begin{cases} -{1\over f^2}\gamma\nabla^\mu\pi & A=\mu \\ \gamma & A=5\end{cases},\ \ \ \ n_A=\begin{cases} -\gamma\nabla_\mu\pi & A=\mu \\ \gamma & A=5\end{cases} \ .
\ee

Using the non-vanishing Christoffel symbols $\Gamma^\lambda_{\mu\nu}=\Gamma^\lambda_{\mu\nu}(g)$, $\Gamma^5_{\mu\nu}=-f f' g_{\mu\nu}$, $\Gamma^\mu_{\nu 5}= \delta^\mu_\nu {f'\over f}$, the extrinsic curvature 
is then
\be 
K_{\mu\nu}=\gamma\left(-\nabla_\mu\nabla_\nu\pi+f f'g_{\mu\nu}+2{f'\over f}\nabla_\mu\pi\nabla_\nu\pi\right) \ .
\ee
Note that when the $4$d coordinates have dimensions of length, $\pi$ has mass dimension $-1$ and $f$ is dimensionless.

The algebra of Killing vectors of $G_{AB}$ contains a natural subalgebra consisting of the Killing vectors
for which $K^5=0$.  This is the subalgebra of Killing vectors that are parallel to the foliation of constant $\rho$ surfaces, and it generates the subgroup of isometries which preserve the foliation.  We choose a basis of this subalgebra and index the basis elements by $i$, 
\be 
K_i^A(X)=\begin{cases} K_i^\mu(x) & A=\mu \\ 0 & A=5\end{cases} \ ,
\ee
where we have written $K_i^\mu(x)$ for the $A=\mu$ components, indicating that these components are independent of $\rho$. To see that this is the case, note that,
for those vectors with $K^5=0$, the $\mu5$ Killing equations (\ref{killingequation}) tell us that $K_i^\mu(x)$ is independent of $\rho$.  Furthermore,  the $\mu\nu$ Killing equations tell us that $K_i^\mu(x)$ is a Killing vector of $g_{\mu\nu}$.   

We now extend our basis of this subalgebra to a basis of the algebra of all Killing vectors by appending a suitably chosen set of linearly independent Killing vectors with non-vanishing $K^5$.  We index these with $I$, so that $(K_i,K_I)$ is a basis of the full algebra of Killing vectors.  From the $55$ component of Killing's equation, we see that $K^5$ must be independent of $\rho$, so we may write $K^5(x)$.  

A general global symmetry transformation thus reads
\be 
\delta_KX^A=a^i K_i^A(X)+a^I K_I^A(X) \ ,
\ee
where $a^i$ and $a^I$ are arbitrary constant coefficients of the transformation.  In the gauge (\ref{physgauge}), the transformations become, from~(\ref{gaugefixsym}),
\be 
\label{specialcasesym} 
(\delta_K+\delta_{g,{\rm comp}})\pi=-a^i K_i^\mu(x)\partial_\mu\pi+a^I K_I^5(x)-a^I K_I^\mu(x,\pi)\partial_\mu\pi \ .
\ee
From this, we see that the $K_i$ symmetries are linearly realized, whereas the $K_I$ are realized nonlinearly.  Thus, the algebra of all Killing vectors is spontaneously broken to the subalgebra of Killing vectors preserving the foliation.

\subsection{\label{maxsymsubsection2} Maximally symmetric cases}

\begin{figure} %  figure placement: here, top, bottom, or page
   \centering
   \includegraphics[width=5.0in]{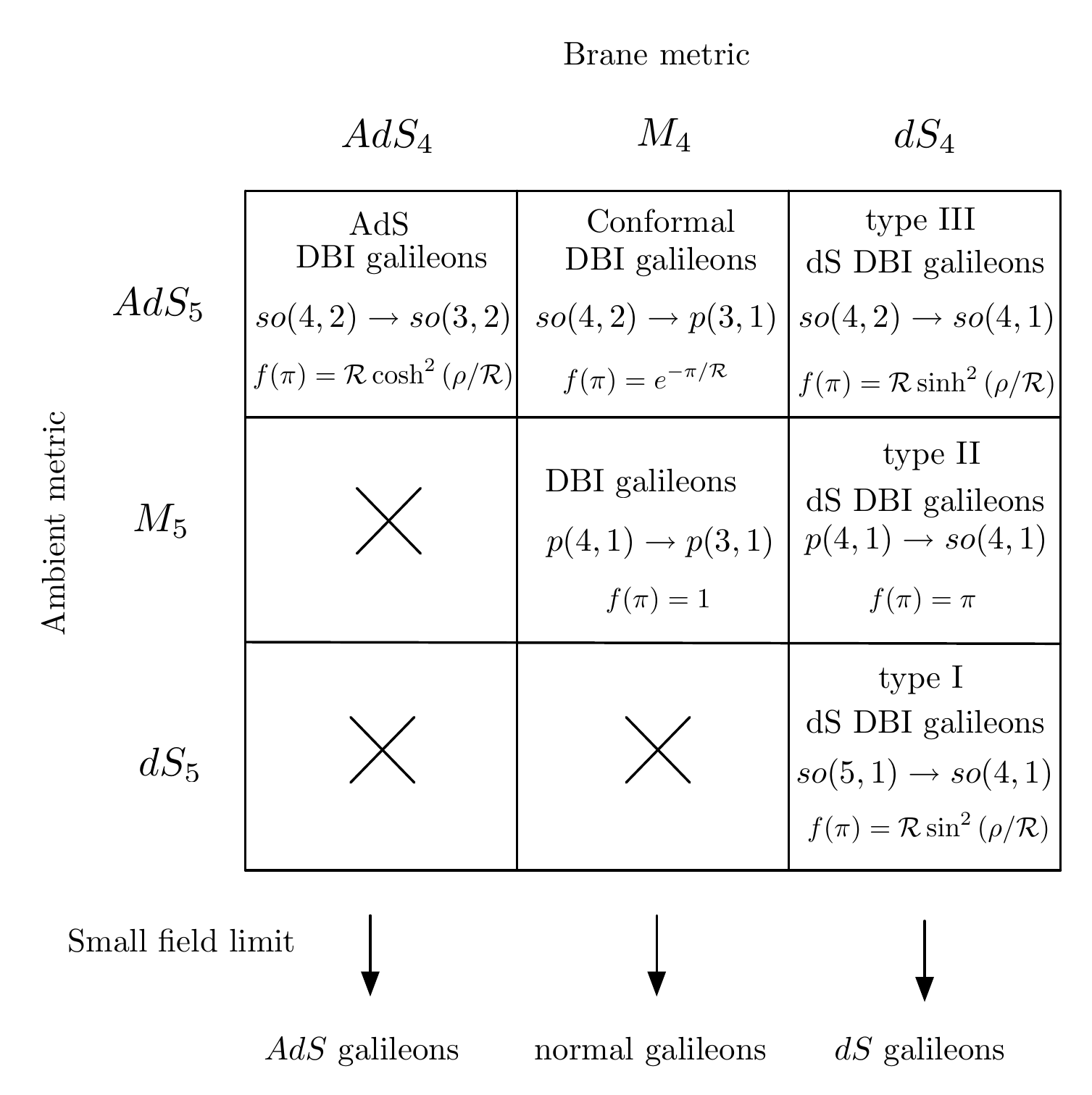}
   \caption{Types of maximally symmetric embedded brane effective field theories, their symmetry breaking patterns, and functions $f(\pi)$.  The relationships
   to the Galileon and DBI theories are also noted.}
   \label{types}
\end{figure}

In this paper, we will focus on the case in which the 5d background metric has 15 global symmetries, the maximal number.  Thus, the bulk is either $5$d anti-de Sitter space $AdS_5$ with isometry algebra $so(4,2)$, 5d de-Sitter space $dS_5$ with isometry algebra $so(5,1)$, or flat 5d Minkowski space $M_5$ with isometry algebra the five dimensional Poincare algebra $p(4,1)$.  In addition, we focus on the case where the brane metric $g_{\mu\nu}$, and hence the extrinsic curvature, are maximally symmetric, so that the unbroken subalgebra has the maximal number of generators, 10.  This means that the leaves of the foliation are either $4$d anti-de Sitter space $AdS_4$ with isometry algebra $so(3,2)$, 4d de-Sitter space $dS_4$ with isometry algebra $so(4,1)$, or flat 4d Minkowski space $M_4$ with isometry algebra the four dimensional Poincare algebra $p(3,1)$.  In fact, there are only 6 such possible foliations of $5$d maximally symmetric spaces by $4$d maximally symmetric time-like slices, such that the metric takes the form~(\ref{metricform}).  Flat $M_5$ can be foliated by flat $M_4$ slices or by $dS_4$ slices; $dS_5$ can be foliated by flat $M_4$ slices, $dS_4$ slices, or $AdS_4$ slices; and $AdS_5$ can only be foliated by $AdS_4$ slices.  Each of these 6 foliations, through the construction leading to~(\ref{gaugefixedaction}), will generate a class of theories living on an $AdS_4$, $M_4$ or $dS_4$ background and having 15 global symmetries  broken to the 10 isometries of the brane.  These possibilities are summarized in Figure \ref{types}.

It should be noted that the missing squares in Figure \ref{types} may be filled in if we are willing to consider a bulk which has more than one time direction\footnote{We thank Sergei Dubovsky for pointing this out.}.  For example, it is possible to embed $AdS_4$ into a five-dimensional Minkowski space with two times (indeed, this is the standard way of constructing $AdS$ spaces).  From the point of view that the bulk is physical, and hence should be thought of as dynamical, these possibilities may be unacceptable on physical grounds.  However, if one thinks of the bulk as merely a mathematical device for constructing novel four-dimensional effective theories, then there is nothing a priori to rule out these possibilities.  In this paper, we focus on those cases in which the bulk has only one time dimension.  The construction in the other cases will, however, follow the same pattern.  

Finally, note that the only invariant data that go into constructing a brane theory are the background metric and the action.  Theories with the same background metric and the same action are isomorphic, regardless of the choice of foliation (which is merely a choice of gauge).  For example, given the same action among the theories listed in Figure \ref{types}, the three that have an $AdS_5$ background, namely the conformal DBI Galileons, the $AdS_4$ DBI Galileons, and the type III $dS_4$ DBI Galileons, are really the same theory.  They are related by choosing a different foliation (gauge), shuffling the background $\pi$ configuration into the background metric.

\section{Actions with second order equations of motion}
\label{sec:ghostfreeactions}
Up until now we have discussed the degrees of freedom and their symmetries, but it is the choice of action that defines the dynamics. A general choice for the function $F$ in~(\ref{gaugefixedaction}) will lead to scalar field equations for $\pi$ which are higher than second order in derivatives.  When this is the case, the scalar will generally propagate extra degrees of freedom which are ghost-like~\cite{Ostrogradski,deUrries:1998bi}.  The presence of such ghosts signifies that either the theory is unstable, or the cutoff must be lowered so as to exclude the ghosts.  Neither of these options is particularly attractive, and so it is desirable to avoid ghosts altogether.  It is the Galileon terms which are special because they lead to equations of at most second order.  Furthermore, as mentioned in the introduction, there can exist regimes in which the Galileon terms dominate over all others, so we will be interested only in these terms.

A key insight of de Rham and Tolley~\cite{deRham:2010eu} is that there are a finite number of actions of the type (\ref{gaugefixedaction}), the Lovelock terms and their boundary terms, that do in fact lead to second order equations for $\pi$ and become the Galileon terms.
The possible
extensions of Einstein gravity which remain second order are given by Lovelock terms~\cite{Lovelock:1971yv}.  These terms are specific combinations of powers of the Riemann tensor which are topological (i.e. total derivatives) in some specific home dimension, but in lower dimensions have the property that equations of motions derived from them are second order.  (For a short summary of some properties of these terms, see Appendix B of~\cite{Hinterbichler:2010xn}.)  The Lovelock terms come with boundary terms.  It is well known that, when a brane is present, bulk gravity described by the Einstein-Hilbert Lagrangian should be supplemented by the Gibbons-Hawking-York boundary term~\cite{Gibbons:1976ue,York:1972sj}
\begin{equation}
\label{e:EHGH}
S = \int_M  d^5 X \ \sqrt{-G}R[G] \ + 2 \int \ d^4 x\sqrt{-\bar g}K \ .
\end{equation}
Similarly, Lovelock gravity in the bulk must be supplemented by brane terms which depend on the intrinsic and extrinsic curvature of the
brane (the so-called Myers terms ~\cite{Myers:1987yn,Miskovic:2007mg}), which are needed in order to make the variational problem for the brane/bulk system well posed~\cite{Dyer:2008hb}.
Of course we are not considering bulk gravity to be dynamical, but the point here is that these boundary terms also yield second order equations of motion for $\pi$ in the construction leading to~(\ref{gaugefixedaction}).

The prescription of~\cite{deRham:2010eu} is then as follows: on the 4-dimensional brane, we may add the first two Lovelock terms, namely the cosmological constant term $\sim \sqrt{-\bar g}$ and the Einstein-Hilbert term $\sim \sqrt{-\bar g}R[\bar g]$.  (The higher Lovelock terms are total derivatives in 4-dimensions.)  We may also add the boundary term corresponding to a bulk Einstein-Hilbert term, $\sqrt{-\bar g}K$, and the boundary term ${\cal L}_{\rm GB}$ corresponding to the Gauss-Bonnet Lovelock invariant $R^2 - 4 R_{\mu\nu} R^{\mu\nu}+ R_{\mu\nu\alpha\beta} R^{\mu\nu\alpha\beta}$ in the bulk.  The zero order cosmological constant Lovelock term in the bulk has no boundary term (although as we will see, we may construct a fifth term, the tadpole term, from it) and the higher order bulk Lovelock terms vanish identically.  Therefore, in total, for a 3-brane there are four possible terms (five including the tadpole) which
lead to second order equations.  These are the terms we focus on.

\subsection{The tadpole term}

As mentioned, there is one term that contains no derivatives of $\pi$ and is not of the form~(\ref{gaugefixedaction}).   This Lagrangian is called the tadpole term, denoted by ${\cal A}(\pi)$.  The value of the tadpole action is the proper 5-volume between some $\rho= {\rm constant}$ surface and the position of the brane,
\be 
\label{tadpoleterm}  
S_1=\int d^4x \int^\pi d\pi' \sqrt{-G}=\int d^4x\sqrt{-g}\int^\pi d\pi' f(\pi')^4,
\ee
so that 
\be
{\cal L}_1=\sqrt{-g}{\cal A}(\pi),\ \ \ \  {\cal A}(\pi)=\int^\pi d\pi' f(\pi')^4.
\ee
Note that ${\cal A}'(\pi)=f(\pi)^4$.

Under a general nonlinear symmetry $\delta_{\rm K}\pi=K^5(x)-K^\mu(x,\pi)\partial_\mu\pi$ of the type~(\ref{specialcasesym}), its change is
\be 
\delta_{K}{\cal L}_1=\sqrt{-g}{\cal A}'(\pi)\delta_{K}\pi=\sqrt{-g}f^4\left(K^5(x)-K^\mu(x,\pi)\partial_\mu\pi\right) \ .
\ee
Using the Killing equation (\ref{killingequation}), it is straightforward to check directly that a general variation of the right-hand side vanishes,
demonstrating that the change in the tadpole term under the symmetry transformation is a total derivative.  Thus the tadpole term has the same symmetries as the other terms.

\subsection{Explicit expressions for the terms}

Including the tadpole term there are thus five terms that lead to second order equations for $\pi$,
\bea   
{\cal L}_1&=&\sqrt{-g}\int^\pi d\pi' f(\pi')^4,\nn\\
{\cal L}_2&=&- \sqrt{-\bar g} \ ,\nn\\
{\cal L}_3&=& \sqrt{-\bar g}K \ ,\nn\\
{\cal L}_4&=& -\sqrt{-\bar g}\bar R \ ,\nn\\
{\cal L}_5&=&{3\over 2}\sqrt{-\bar g} {\cal K}_{\rm GB} \ ,
\label{ghostfreegenterms}
\eea
where the explicit form of the Gauss-Bonnet boundary term is
\be 
{\cal K}_{\rm GB}=-{1\over3}K^3+K_{\mu\nu}^2K-{2\over 3}K_{\mu\nu}^3-2\left(\bar R_{\mu\nu}-\half \bar R \bar g_{\mu\nu}\right)K^{\mu\nu} \ .
\ee  
Indices are raised and traces are taken with $\bar g^{\mu\nu}$.  At this stage, each of these terms would appear in a general Lagrangian with an arbitrary coefficient. As we will see later, requiring stability will, however, force certain choices on us in specific examples.

En route to presenting specific examples of our new theories, we now evaluate these terms on the special case metric~(\ref{metricform}).  We make use of formulae catalogued in Appendix \ref{appendix1}.  Our strategy is to collect coefficients of $f''$, $f'$, $f'^2$ and $f'^3$, eliminate everywhere $(\partial\pi)^2$ in favor of $\gamma={1\over \sqrt{1+{1\over f^2}(\partial\pi)^2}}$, and then to group like terms by powers of $\gamma$. A lengthy calculation yields
\bea   
{\cal L}_1&=&\sqrt{-g}\int^\pi d\pi' f(\pi')^4,\nn\\
{\cal L}_2&=&-\sqrt{-g}f^4\sqrt{1+{1\over f^2}(\partial\pi)^2},\nn\\
{\cal L}_3&=&\sqrt{-g}\left[f^3f'(5-\gamma^2)-f^2[\Pi]+\gamma^2[\pi^3]\right],\nn\\
{\cal L}_4&=& -\sqrt{-g}\left\{{1\over\gamma}f^2R-2{\gamma}R_{\mu\nu}\nabla^\mu\pi\nabla^\nu\pi\right. \nn\\
&&+\gamma\left[[\Pi]^2-[\Pi^2]+2{\gamma^2\over f^2}\left(-[\Pi][\pi^3]+[\pi^4]\right)\right]+6{f^3f''\over \gamma}\left(-1+\gamma^2\right) \nn\\
&&\left.+2\gamma ff'\left[-4[\Pi]+{\gamma^2\over f^2}\left(f^2[\Pi]+4[\pi^3]\right)\right]-6{f^2f'^2\over \gamma}\left(1-2\gamma^2+\gamma^4\right) \right\},\nn\\
{\cal L}_5&=&{3\over 2}\sqrt{-g}\left\{ R\left[3ff'-[\Pi]+{\gamma^2\over f^2}\left(-f^3f'+[\pi^3]\right)\right]-2{\gamma^2\over f^2}R^{\mu\nu\alpha\beta}\nabla_\mu\pi\nabla_\alpha\pi\Pi_{\nu\beta} \right.\nn \\
&& +2R^{\mu\nu}\left[\Pi_{\mu\nu}+{\gamma^2\over f^2}\left((-3ff'+[\Pi])\nabla_\mu\pi\nabla_\nu\pi-2\Pi_{\alpha(\mu}\nabla_{\nu)}\pi\nabla^\alpha\pi\right)\right] \nn\\
&&-{\gamma^2\over f^2}\left[{2\over 3}\left([\Pi]^3-3[\Pi][\Pi^2]+2[\Pi^3]\right)+2{\gamma^2\over f^2}\left(-[\pi^3]([\Pi]^2-[\Pi^2])+2[\Pi][\pi^4]-2[\pi^5]\right)\right]\nn\\
&& +4ff''\left[-3ff'+[\Pi]+{\gamma^2\over f^2}\left(3f^3f'-f^2[\Pi]-[\pi^3]\right)\right]-2ff'^3\left(9-11\gamma^2+6\gamma^4\right)\nn\\
&&+2f'^2\left[[\Pi]-{\gamma^2\over f^2}\left(8f^2[\Pi]+[\pi^3]\right)+2{\gamma^4\over f^2}\left(2f^2[\Pi]+5[\pi^3]\right)\right] \nn\\
&& \left.+2\gamma^2{f'\over f}\left[3\left([\Pi]^2-[\Pi^2]\right)-{\gamma^2\over f^2}\left(f^2([\Pi]^2-[\Pi^2])+6([\Pi][\pi^3]-[\pi^4])\right)\right] \right\} \ . \nn\\
\label{generalterms}
\eea
The quantities $[\Pi^n]$ and $[\pi^n]$ are various contractions of derivatives of the $\pi$ field, and the notation is explained in the conventions at the end of Section \ref{introduction}.  In these expressions, all curvatures are those of the metric $g_{\mu\nu}$, and all derivatives are covariant derivatives with respect to $g_{\mu\nu}$.  We point out that no integrations by parts have been performed in obtaining these expressions.  

The equations of motion derived from any of these five terms will 
contain no more than two derivatives on each field, ensuring that no extra degrees of freedom propagate around any background.  After suitable integrations by parts, these actions should therefore conform to the general structure presented in \cite{Deffayet:2011gz} for actions of a single scalar with second order equations (see also the Euler hierarchy constructions \cite{Fairlie:1991qe,Fairlie:1992nb,Fairlie:1992yy,Fairlie:2011md}).  In the above construction, however, we can immediately identify the nonlinear symmetries by reading them off from the isometries of the bulk.

Finally, we note that by keeping the metric $g_{\mu\nu}$ in (\ref{metricform}) arbitrary rather than fixing it to the foliation, we can automatically obtain the covariantizaton of these various Galileon actions, including the non-minimal curvature terms required to keep the equations of motion second order, the same terms obtained by purely 4-d methods in \cite{Deffayet:2009mn,Deffayet:2009wt,Deffayet:2011gz}.  Of course, this in general ruins the symmetries we are interested in considering.  But from this point of view, we can see exactly when such symmetries will be present.  The symmetries will only be present if the $g_{\mu\nu}$ which is used to covariantly couple is such that the full metric (\ref{metricform}) has isometries.

%%%%%%%%%%%%%%%%%%%%
\section{Maximally Symmetric Examples}

We now proceed to construct explicitly the maximally symmetric examples catalogued in Section \ref{maxsymsubsection2} and Figure \ref{types}.  The construction starts by finding coordinates which are adapted to the desired foliation, so that the metric in the bulk takes the form (\ref{metricform}), allowing us to read off the function $f(\pi)$.  Plugging into (\ref{generalterms}) then gives us the explicit Lagrangians.  To find the form of the global symmetries, we must write the explicit Killing vectors in the bulk, and identify those which are parallel and not parallel to the foliation.  We may then read off the symmetries from (\ref{specialcasesym}).  

The construction for each case is similar, and some of the results are related by analytic continuation, but there are enough differences in the forms of the embeddings and the Killing vectors that we thought it worthwhile to display each case explicitly.  The reader interested only in a given case may skip directly to it.

\subsection{A Minkowski brane in a Minkowski bulk: $M_4$ in $M_5$ -- DBI Galileons}

Choosing cartesian coordinates $(x^\mu,\rho)$ on $M_5$, the foliation of $M_5$ by $M_4$ is simply given by $\rho={\rm constant}$ slices, and the metric takes the form
\be 
ds^2=(d\rho)^2+\eta_{\mu\nu}dx^\mu dx^\nu \ .
\ee  
Comparing this to~(\ref{metricform}), we obtain
\be
{f(\pi)=1,\ \ \ g_{\mu\nu}=\eta_{\mu\nu},}
\ee
and the terms~(\ref{generalterms}) become (again, without integration by parts)
\begin{align}
\mathcal{L}_{1}&=\pi ,\nn \\
\mathcal{L}_{2}&=-\sqrt{1+(\partial \pi)^{2}} \ , \nn \\
\mathcal{L}_{3}&=-\left [\Pi\right ]+\gamma^{2}\left [\pi^3\right ] \ ,\nn \\
\mathcal{L}_{4}& =-\gamma \left (\left [\Pi \right ]^{2} -\left [\Pi^{2}\right ]\right )-2\gamma^{3}\left (\left [\pi^{4}\right ]-\left [\Pi\right ]\left [\pi^3\right ]\right ) \ ,\nn \\
\mathcal{L}_{5}& =-\gamma^{2}\left (\left [\Pi\right ]^{3}+2\left [\Pi^{3}\right ]-3\left [\Pi\right ]\left [\Pi^{2}\right ]\right )-\gamma^{4}\left (6\left [\Pi\right ]\left [\pi^{4}\right ]-6\left [\pi^{5}\right ]-3\left (\left [\Pi\right ]^{2}-\left [\Pi^{2}\right ]\right )\left [\pi^3\right ]\right ) \ ,\nn \\
\label{DBIGalileonterms}
\end{align}
where $\gamma={1\over \sqrt{1+(\partial\pi)^2}}$.  
These are the DBI Galileon terms, first written down in \cite{deRham:2010eu} and further studied in \cite{Goon:2010xh}. 

\subsubsection{Killing vectors and symmetries}

The Killing vectors of $5$d Minkowski space are the 10 boosts $L_{AB}=X_A\partial_B-X_B\partial_A$, and the 5 translations $P_A=-\partial_A$.  The 6 boosts $J_{\mu\nu}$ and the 4 translations $P_\mu$ are parallel to the foliation and form the unbroken $p(3,1)$ symmetries of $M_{4}$.  The 5 broken generators are
\bea 
&&K\equiv-P_5=\partial_\rho, \\
&& K_\mu\equiv L_{\mu 5}=x_\mu\partial_\rho-\rho\partial_\mu \ .
\eea
Using the relation $\delta_K\pi=K^5(x)-K^\mu(x,\pi)\partial_\mu\pi$ from~(\ref{specialcasesym}), we obtain the transformation rules
\bea 
&&\delta\pi=1, \nn \\
&& \delta_\mu \pi=x_\mu+\pi\partial_\mu\pi \ , \label{DBIGalileontrans}
\eea
under which the terms (\ref{DBIGalileonterms}) are each invariant up to a total derivative.  The symmetry breaking pattern is
\be 
p(4,1)\rightarrow p(3,1) \ .
\ee

%%%%%%%%%%%%%%%%%%%%
\subsection{A Minkowski brane in an anti-de Sitter bulk: $M_4$ in $AdS_5$ -- Conformal Galileons\label{MinAdSsection}}

In this section, indices $\Ac,\Bc,\cdots$ run over six values $0,1,2,3,4,5$ and $Y^\Ac$ are cartesian coordinates in an ambient $6$d two-time Minkowski space with metric $\eta_{\Ac\Bc}={\rm diag}(-1,-1,1,1,1,1)$, which we call $M_{4,2}$.  

Five dimensional anti-de Sitter space $AdS_5$ (more precisely, a quotient thereof) can be described as the subset of points $(Y^0,Y^1,Y^2\ldots,Y^{5})\in M_{4,2}$ in the hyperbola of one sheet satisfying 
\be 
\eta_{\Ac\Bc}Y^\Ac Y^\Bc=-(Y^0)^2-(Y^1)^2+(Y^2)^2+\cdots+(Y^{5})^2=-{\cal R}^2 \ ,
\ee
with ${\cal R}>0$ the radius of curvature of $AdS_5$, and where the metric is induced from the flat metric on $M_{4,2}$.  This space is not simply connected, but its universal cover is $AdS_5$.  The scalar curvature $R$ and cosmological constant $\Lambda$ are given by $R=-{20\over {\cal R}^2},\ \Lambda=-{6\over{\cal R}^2}.$

We use Poincare coordinates $(\rho, x^\mu)$ on $AdS_5$ which cover the region $Y^0+Y^2>0$,
\bea 
\nn Y^0&=&\rc \cosh\left(\rho\over\rc\right)+{1\over 2\rc}e^{-\rho/\rc}x^2 \ ,\\ \nn
Y^1&=&e^{-\rho/\rc}x^0 \ ,\\ \nn
Y^2&=&-\rc \sinh\left(\rho\over\rc\right)-{1\over 2\rc}e^{-\rho/\rc}x^2 \ ,\\
Y^{i+2}&=&e^{-\rho/\rc}x^i \ ,\ \ i=1,2,3 \ ,
\eea
where $x^2\equiv\eta_{\mu\nu}x^\mu x^\nu$, and $\eta_{\mu\nu}={\rm diag}(-1,1,1,1)$ is the Minkowski 4-metric.  The coordinates $u$ and $x^\mu$ all take the range $(-\infty,\infty)$.  Lines of constant $\rho$ foliate the Poincare patch of $AdS_5$ with Minkowski $M_4$ time-like slices, given by intersecting the planes $Y^0+Y^2={\rm constant}$ with the hyperbola.

The induced metric is
\be 
ds^2=d\rho^2+e^{-2\rho/\rc}\eta_{\mu\nu}dx^\mu dx^\nu \ .
\ee
Comparing this with~(\ref{metricform}) we obtain    
\be
{f(\pi)=e^{-\pi/\rc},\ \ \ \ g_{\mu\nu}=\eta_{\mu\nu}} \ ,
\ee  
and the terms~(\ref{generalterms}) become (without integration by parts)
\bea 
{\cal L}_1&=& -{\rc\over 4}e^{-4\pi/ \rc} \ ,\nn\\
{\cal L}_2&=& -e^{-4\pi/ \rc}\sqrt{1+e^{2\pi/ \rc} (\partial\pi)^2} \ ,\nn \\
{\cal L}_3&=& \gamma^2[\pi^3]-e^{-2\pi/ \rc}[\Pi]+{1\over \rc}e^{-4\pi/ \rc}(\gamma^2-5) \ ,\nn \\
{\cal L}_4&=&
-\gamma([\Pi]^2-[\Pi^2])-2\gamma^3 e^{2\pi/\rc}([\pi^4]-[\Pi][\pi^3])\nn\\
&&+\frac{6}{\rc ^2}e^{-4\pi/\rc}{1\over \gamma}\(2-3\gamma^2+\gamma^4\)+\frac 8 \rc\gamma^3 [\pi^3]
-\frac 2 \rc e^{-2\pi/\rc}\gamma\(4-\gamma^2\)[\Pi]
 \, ,\nn \\
{\cal L}_5&=&
-\gamma^2 e^{2\pi/ \rc}\([\Pi]^3-3[\Pi][\Pi^2]+2[\Pi^3]\) \nn \\ 
\hspace{-5pt}&&\hspace{-5pt}
-3 \gamma^4e^{4\pi/ \rc}\left[2([\Pi][\pi^4]-[\pi^5])-([\Pi]^2-[\Pi^2])[\pi^3]\right]\nn\\
\hspace{-5pt}&&\hspace{-5pt} +\frac{18}{ \rc}e^{2\pi/ \rc}\gamma^4([\Pi][\pi^3]-[\pi^4])-\frac 3  \rc {\gamma^2}(3-\gamma^2)([\Pi]^2-[\Pi^2])\nn\\
\hspace{-5pt}&&\hspace{-5pt}-\frac{3}{ \rc^2}{ \gamma^2}(3-10\gamma^2)[\pi^3]
-\frac{3}{ \rc^2}e^{-2\pi/ \rc}(-3+10\gamma^2-4\gamma^4)[\Pi]\nn \hspace{-100pt}\nn \\
\hspace{-5pt}&&\hspace{-5pt}+\frac{3}{ \rc^3}e^{-4\pi/ \rc}(15-17\gamma^2+6\gamma^4)
\ , \nn \\ \label{conformalDBIGalileonterms}
\eea
where 
\be
\gamma={1\over \sqrt{1+e^{2\pi/ \rc} (\partial\pi)^2}} \ .
\ee
These are the conformal DBI Galileons, first written down in \cite{deRham:2010eu}.

\subsubsection{Killing vectors and symmetries}

The 15 Lorentz generators of $M_{4,2}$; $M_{ \Ac  \Bc }=Y_ \Ac \partialb_ \Bc -Y_ \Bc \partialb_ \Ac $ (here $\partialb_ \Ac $ are the coordinate basis vectors in the ambient space $M_{4,2}$, and indices are lowered with the $M_{4,2}$ flat metric $\eta_{ \Ac  \Bc }$) are all tangent to the $AdS_5$ hyperboloid, and become the 15 isometries of the $so(4,2)$ isometry algebra of $AdS_5$.  Of these, 10 have no $\partial_\rho$ components and are parallel to the $M_4$ foliation.  These form the unbroken $p(3,1)$ isometry algebra of the $M_4$ slices.  

First we have
\bea 
Y^{i+2}\partialb_1+Y^1\partialb_{i+2}&\rightarrow &x^i\partial_0+x^0\partial_i ,\ \ \ i=1,2,3,\\
Y^{i+2}\partialb_{j+2}-Y^{j+2}\partialb_{i+2}&\rightarrow & x_i\partial_j-x_j\partial_i, \ \ \ i,j=1,2,3,
\eea
which taken together are the 6 Lorentz transformations $L_{\mu\nu}=x_\mu\partial_\nu-x_\nu\partial_\mu$ of the $x^{\mu}$.

For the remaining 4, we focus on
\bea 
-Y^{1}\partialb_0+Y^0\partialb_{1}&\rightarrow &x^0\partial_\rho+\left[{\rc\over 2}\left(1+e^{2\rho\over \rc}\right)+{1\over 2\rc}x^2\right]\partial_0+{1\over \rc}x^0x^\mu\partial_\mu \ ,\nn \\
 -Y^{i+2}\partialb_0-Y^0\partialb_{i+2}&\rightarrow &x^i\partial_\rho+\left[-{\rc\over 2}\left(1+e^{2\rho\over \rc}\right)-{1\over 2\rc}x^2\right]\partial_i+{1\over \rc}x^ix^\mu\partial_\mu \ ,\ \ \ \ i=1,2,3 \ ,\nn \\
  -Y^{2}\partialb_1-Y^1\partialb_{2}&\rightarrow &x^0\partial_\rho+\left[-{\rc\over 2}\left(1-e^{2\rho\over \rc}\right)+{1\over2 \rc}x^2\right]\partial_0+{1\over \rc}x^0x^\mu\partial_\mu \ ,\nn \\
 -Y^{i+2}\partialb_2+Y^2\partialb_{i+2}&\rightarrow &x^i\partial_\rho+\left[{\rc\over 2}\left(1-e^{2\rho\over \rc}\right)-{1\over2 \rc}x^2\right]\partial_i+{1\over \rc}x^ix^\mu\partial_\mu \ ,\ \ \ \ i=1,2,3  \ ,
\eea
which may be grouped as
\bea 
 V_\mu &= &x_\mu\partial_\rho+\left[-{\rc\over 2}\left(1+e^{2\rho\over \rc}\right)-{1\over 2\rc}x^2\right]\partial_\mu+{1\over \rc}x_\mu x^\nu\partial_\nu \ ,\ \ \ \ \mu=0,1,2,3\nn \\
 V'_\mu &= &x_\mu\partial_\rho+\left[{\rc\over 2}\left(1-e^{2\rho\over \rc}\right)-{1\over2 \rc}x^2\right]\partial_\mu+{1\over \rc}x_\mu x^\nu\partial_\nu,\ \ \ \ \mu=0,1,2,3 \ . \nn\\
\eea
If we now take the following linear combinations,
\bea 
P_\mu &= & {1\over \rc}(V_\mu-V'_\mu)=-\partial_\mu \ ,\\
K_\mu &= &(V_\mu+V'_\mu)=2x_\mu\partial_\rho-\left[{\rc}e^{2\rho\over \rc}+{1\over \rc}x^2\right]\partial_\mu+{2\over \rc}x_\mu x^\nu\partial_\nu \ ,
\eea
the $P_\mu$ are the translations on the $x^{\mu}$, the remaining 4 unbroken vectors.  

The $K_\mu$ are broken generators and, in addition, there is one more broken vector, 
\be 
-Y^2\partialb_0-Y^0\partialb_2=\rc\partial_\rho+x^\mu\partial_\mu \ .
\ee
Using the relation $\delta_K\pi=K^5(x)-K^\mu(x,\pi)\partial_\mu\pi$ from~(\ref{specialcasesym}), we obtain the transformation rules for the $\pi$ field from this and from the $K_\mu$ as
\bea 
\delta\pi&=&\rc-x^\mu\partial_\mu\pi,\nn\\
\delta_\mu\pi&=&2x_\mu+\left[{\rc}e^{2\pi\over \rc}+{1\over \rc}x^2\right]\partial_\mu\pi-{2\over \rc}x_\mu x^\nu\partial_\nu\pi \ . \label{MinAdStrans}
\eea
The terms~(\ref{conformalDBIGalileonterms}) are each invariant up to a total derivative under these transformations, and the symmetry breaking pattern is
\be 
so(4,2)\rightarrow p(3,1) \ .
\ee

%%%%%%%%%%%%%%%%%%%%
\subsection{A de Sitter brane in a Minkowski bulk: $dS_4$ in $M_5$}

We describe the Minkowski bulk with the usual metric in cartesian coordinates
\be
{ds^2=\eta_{AB}X^AX^B=-(dX^0)^2+(dX^1)^2+(dX^2)^2+(dX^3)^2+(dX^4)^2} \ .
\ee
The region $\eta_{AB}X^AX^B>0$ (i.e. outside the lightcone) can be foliated by de Sitter slices.  To see this, we use Rindler coordinates which cover this region,
\bea 
\nn X^0&=&r\sinh\tau, \\ \nn
X^1&=&\rho \cosh \tau\ \cos\theta_1 \ , \\ \nn
X^2&=& \rho\cosh \tau\ \sin\theta_1\cos\theta_2 \ , \\ \nn
X^3&=& \rho\cosh \tau\ \sin\theta_1\sin\theta_2\cos\theta_3 \ , \\
X^4&=& \rho\cosh \tau\ \sin\theta_1\sin\theta_2\sin\theta_3 \ ,
\eea
where $\rho\in(0,\infty)$, $\tau\in(-\infty,\infty)$, and the $\theta_i$ ($i=1,2,3$) parametrize a $3$ sphere. The metric in Rindler coordinates is then
\be
\label{rindlermetric}
{ds^2=d \rho ^2+ \rho ^2\left[-d\tau^2+\cosh^2\tau \ d\Omega^2_{(3)}\right]} \ .
\ee
This metric is $ds^2=d\rho^2+\rho^2 ds^2_{dS_{4}},$ where $ds^2_{dS_{4}}$ is the global metric on a unit radius $4$d de Sitter space.  The foliation by $dS^{4}$ thus corresponds to $\rho= {\rm constant}$ surfaces (or to $-(X^0)^2+(X^i)^2= {\rm constant} >0$ in cartesian coordinates).  

Comparing this with~(\ref{metricform}), we obtain 
\be
{f(\pi)=\pi,}\ \ \ g_{\mu\nu}=g_{\mu\nu}^{(dS_4)} \ ,
\ee
and the terms~(\ref{generalterms}) become (without any integrations by parts)
\bea
{\cal L}_1&=&{1\over 5}\sqrt{-g}{\pi^5} \ ,\nn\\
{\cal L}_2&=&-\sqrt{-g}\pi^4\sqrt{1+{1\over \pi^2}(\partial\pi)^2} \ ,\nn \\
{\cal L}_3&=&\sqrt{-g}\left[\pi^3(5-\gamma^2)-\pi^2[\Pi]+\gamma^2[\pi^3]\right] \ ,\nn \\
{\cal L}_4&=&\sqrt{-g}\ \gamma\left[-[\Pi]^2+[\Pi^2]+8\pi[\Pi]-18\pi^2-2{\gamma^2\over \pi^2}\left([\pi^4]+4\pi[\pi^3]-3\pi^4-[\Pi][\pi^3]+\pi^3[\Pi]\right)\right] \ ,\nn \\
{\cal L}_5&=&\sqrt{-g}\ {\gamma^2\over \pi^2}\Bigg[ -[\Pi]^3+3[\Pi][\Pi^2]-2[\Pi^3]+9\pi([\Pi]^2-[\Pi^2])+42\pi^3-30\pi^2[\Pi]           \nn\\
&&+3{\gamma^2\over \pi^2}\left(([\Pi]^2-[\Pi^2])[\pi^3]+2[\pi^5]+6\pi[\pi^4]+10\pi^2[\pi^3]-\pi^3([\Pi]^2-[\Pi^2]) \right. \nn\\
&&\left. -6\pi^5-2[\Pi]([\pi^4]+3\pi[\pi^3]-2\pi^4)]\right)\Bigg] \ , \label{dSinMDBI}
\eea
where the background metric and covariant derivatives are those of unit-radius $4$d de Sitter space, and
\be 
\gamma={1\over \sqrt{1+{1\over \pi^2}(\partial\pi)^2}} \ .
\ee

Note that, since we have chosen the $4$d space to be a unit-radius $dS_4$ with dimensionless coordinates, $\pi$ and $f$ have mass dimension $-1$.  In evaluating (\ref{dSinMDBI}), we have used that the scalar curvature and cosmological constant of this space are $R=12$ and $\Lambda=3$ respectively, and used the relations $R_{\mu\nu\alpha\beta}={R\over 12}\(g_{\mu\alpha}g_{\nu\beta}-g_{\mu\beta}g_{\nu\alpha}\)$, and $R_{\mu\nu}={R\over 4}g_{\mu\nu}$, valid for a maximally symmetric space.  It is possible, of course, to rescale the coordinates, canonically normalize the field, and/or rescale $f$ to bring these quantities to their usual dimensions.  Given a suitable combinations of these Lagrangians so that a constant field $\pi(x)=\pi_0={\rm constant}$ is a solution to the equations of motion, $\pi_0$ sets the radius of the de Sitter brane in its ground state.  

We call these Type II de Sitter DBI Galileons (see Figure \ref{types}), and they are our first example of a Galileon that lives on curved space yet still retains the same number of shift-like symmetries as their flat space counterparts.

\subsubsection{Killing vectors and symmetries}

The 10 Lorentz transformations of $M_5$ are parallel to the de Sitter slices and become the unbroken $so(4,1)$ isometries of $dS_4$.  The 5 translations are not parallel and will be nonlinearly realized.

With a future application to cosmology in mind, we will calculate the transformation laws explicitly using conformal inflationary coordinates $(u,y^i)$ on the de Sitter slices, even though these coordinates only cover half of each de Sitter slice.  The embedding becomes
\bea 
X^0&=&{\rho\over 2u}\left(1-u^2+y^2\right) \ ,\nn \\
X^1&=&{\rho\over 2u}\left(1+u^2-y^2\right) \ ,\nn \\
X^{i+1}&=&{\rho y^i\over u},\ \ \ i=1,2,3 \ ,
\label{dsMinflationaryembedding}
\eea
where $y^2\equiv \delta_{ij}y^iy^j$, and the coordinate ranges are $\rho\in (0,\infty)$, $u\in(0,\infty)$, $y^i\in(-\infty,\infty)$.  The metric takes the form
\be 
ds^2=d\rho^2+\rho^2\left[{1\over u^2}\left(-du^2+dy^2\right)\right] \ ,
\ee
so that the $dS_4$ slices have conformal inflationary coordinates, with $u$ the conformal time.

We are interested in the form of the nonlinear symmetries stemming from the broken translation generators of $M_5$. In the coordinates~(\ref{dsMinflationaryembedding}), the broken Killing vectors $\partialb_A$ are
\bea 
\partialb_0&=&{1\over 2u}\left(-1+u^2-y^2\right)\partial_\rho-{1\over 2\rho}\left(1+u^2+y^2\right)\partial_u-{u\over \rho}y^i\partial_i \ ,\\
 \partialb_1&=&{1\over 2u}\left(1+u^2-y^2\right)\partial_\rho-{1\over 2\rho}\left(-1+u^2+y^2\right)\partial_u-{u\over \rho}y^i\partial_i \ ,\\
 \partialb_{i}&=&{y_i\over u}\partial_\rho+{y_i\over\rho}\partial_u+{u\over r}\partial_i,\ \ i=1,2,3 \ .
\eea
Taking the following linear combinations
\bea 
K_+&=&\partialb_0+\partialb_1={1\over u}\left(u^2-y^2\right)\partial_\rho-{1\over \rho}\left(u^2+y^2\right)\partial_u-{2u\over \rho}y^i\partial_i \ ,\\
K_-&=&\partialb_0-\partialb_1=-{1\over u}\partial_\rho-{1\over \rho}\partial_u \ ,\\
K_i&=& \partialb_{i}={y_i\over u}\partial_\rho+{y_i\over\rho}\partial_u+{u\over \rho}\partial_i \ ,
\eea
and using the relation $\delta_K\pi=K^5(x)-K^\mu(x,\pi)\partial_\mu\pi$ from~(\ref{specialcasesym}), we then obtain the transformation rules
\bea 
\delta_+\pi&=&{1\over u}\left(u^2-y^2\right)+{1\over \pi}\left(u^2+y^2\right)\pi'+{2u\over \pi}y^i\partial_i\pi \ ,\nn\\
\delta_-\pi&=&-{1\over u}+{1\over \pi}\pi' \ ,\nn \\
\delta_i\pi &=& {y_i\over u}-{y_i\over\pi}\pi'-{u\over \pi}\partial_i\pi \ ,
\label{dSinMtrans}
\eea
where $\pi'\equiv \partial_u\pi$.

The terms~(\ref{dSinMDBI}) are each invariant up to a total derivative under these transformations, and the symmetry breaking pattern is
\be 
p(4,1)\rightarrow so(4,1) \ .
\ee

%%%%%%%%%%%%%%%%%%%%
\subsection{A de Sitter brane in a de Sitter bulk: $dS_4$ in $dS_5$}
In this section, indices $\Ac,\Bc,\cdots$ run over six values $0,1,2,3,4,5$ and $Y^\Ac$ are coordinates in an ambient $6$d Minkowski space with metric $\eta_{\Ac\Bc}={\rm diag}(-1,1,1,1,1,1)$, which we call $M_{6}$.  

Five-dimensional de Sitter space $dS_5$ can be described as the subset of points $(Y^0,Y^1,Y^2\ldots,Y^{5})\in M_{6}$ in the hyperbola of one sheet satisfying 
\be 
\eta_{\Ac\Bc}Y^\Ac Y^\Bc=-(Y^0)^2+(Y^1)^2+(Y^2)^2+\cdots+(Y^{5})^2={\cal R}^2 \ ,
\ee
with the metric induced from the metric on $M_{6}$, for some constant ${\cal R}>0$, the radius of curvature of the $dS_5$. The scalar curvature $R$ and cosmological constant $\Lambda$ are given by $R=20/ {\cal R}^2$ and $\Lambda=6/{\cal R}^2$, respectively.

We use coordinates in which the constant $\rho$ surfaces are the intersections of the planes $Y^1={\rm constant}$ with the hyperbola, and are themselves four-dimensional de Sitter spaces $dS^{4}$,
\bea 
Y^0&=&{\cal R}\sin\rho\sinh\tau \ , \\
Y^1&=&{\cal R}\cos \rho \ , \\
Y^2&=&{\cal R}\cosh\tau\sin\rho\ \cos\theta_1 \ ,\\
Y^3&=&{\cal R}\cosh\tau\sin\rho\ \sin\theta_1\cos\theta_2 \ ,\\
Y^4&=&{\cal R}\cosh\tau\sin\rho\ \sin\theta_1\sin\theta_2\cos\theta_3 \ ,\\
Y^5&=&{\cal R}\cosh\tau\sin\rho\ \sin\theta_1\sin\theta_2\sin\theta_3 \ .\\
\eea
Here $\tau\in(-\infty,\infty)$, $\rho\in(0,\pi)$ 
and $\theta_i$, $i=1,2,3$ parametrize a $3$-sphere. These coordinates cover the region $0<Y^1<{\cal R}$, $0<Y^2<{\cal R}$. 

The metric is
\be
\label{hyperbolicdesitterinside}
dx^2={\cal R}^2\left[d\rho^2+\sin^2\rho\left(-d\tau^2+\cosh^2\tau\ d\Omega_{(3)}\right)\right] \ .
\ee
Scaling $\rho$ so that it lies in the range $(0,\pi \rc)$, the metric becomes $ds^2=d\rho^2+\rc^2\sin^2\(\rho\over \rc\) ds^2_{dS_{4}}$, where $ds^2_{dS_{4}}$ is the global metric on a four-dimensional de Sitter space $dS_4$ of unit radius.  The foliation by $dS^{4}$ thus corresponds to $\rho={\rm constant}$ surfaces.  These slices are given by intersecting the planes $Y^1={\rm constant}$ with the hyperbola, for values $0<Y^1<{\cal R}$.  (By taking $\rho<0$ we cover instead $-{\cal R}<Y^2<0$.  This is the maximum extent to which we may extend the foliation.)

Comparing this with~(\ref{metricform}), we obtain 
\be
{f(\pi)=\rc\sin (\pi/{\cal R}), \ \ \ g_{\mu\nu}=g_{\mu\nu}^{(dS_4)}} \ ,
\ee
and the terms~(\ref{generalterms}) become (using no integrations by parts)
\bea
{\cal L}_1&=& \sqrt{-g}{\rc^4\over 32}\( 12\ \pi-8\rc \sin\(2\pi\over \rc\) +\rc \sin\(4\pi\over \rc\)\) \ ,\\
{\cal L}_2&=&-\sqrt{-g}{\rc^4\over \gamma} \sin^4\(\pi\over \rc\) \ ,\\
{\cal L}_3&=&\sqrt{-g}\left[\gamma^2 [\pi^3]-\rc^2[\Pi] \sin^2\(\pi\over \rc\)+\rc^3(5-\gamma^2) \sin^3\(\pi\over \rc\) \cos\(\pi\over \rc\)\right] \ , \\
{\cal L}_4&=&\sqrt{-g}\Bigg[ {2\gamma^3\over \rc^2}\([\Pi][\pi^3]-[\pi^4]\)\csc^2\pir-{\gamma}\([\Pi]^2-[\Pi^2]+{8\gamma^2\over \rc}[\pi^3]\cot\pir\) \\
&&+\rc\gamma(4-\gamma^2)[\Pi]\sin\pirt+{3\rc^2\over \gamma}\sin^2\pir \(-2-3\gamma^2+\gamma^4+(2-3\gamma^2+\gamma^4)\cos\pirt\)\Bigg] \ , \nn \\
{\cal L}_5&=&\sqrt{-g}\Bigg[ {3\gamma^4\over \rc^4}\(2([\pi^5]-[\Pi][\pi^4])+[\pi^3]([\Pi]^2-[\Pi^2])\)\csc^4\pir \\ 
&&-{18\gamma^4\over \rc^3}\([\Pi][\pi^3]-[\pi^4]\)\csc^2\pir\cot\pir \nn\\
&&-{\gamma^2\over \rc^2}\csc^2\pir\([\Pi]^3-3[\Pi][\Pi^2]+2[\Pi^3]-{3\over 2}(3+10\gamma^2)[\pi^3]+{3\over 2}(3-10\gamma^2)[\pi^3]\cos\pirt \) \nn \\
&&+{3\gamma^2\over \rc}(3-\gamma^2)([\Pi]^2-[\Pi^2])\cot\pir+{3\over 2}[\Pi] \(-3-10\gamma^2+4\gamma^4+(3-10\gamma^2+4\gamma^4)\cos\pirt\) \nn \\
&& -{3\rc \over 4} \(-15-11\gamma^2+6\gamma^4+(15-17\gamma^2+6\gamma^4)\cos\pirt\) \sin\pirt \Bigg] \ , 
\label{dSindSDBI} 
\eea
where the background metric and covariant derivatives are those of the unit-radius $4$d de Sitter space, and
\be 
\gamma={1\over \sqrt{1+{(\partial\pi)^2\over \rc^2 \sin^2\(\pi\over \rc\)}}} \ .
\ee

Since we have chosen the $4$d space to have unit radius in dimensionless coordinates, $\pi$ and $f$ have mass dimension $-1$.  In evaluating~(\ref{dSinMDBI}), we have used that fact that the scalar curvature and cosmological constant of this space are $R=12$ and 
$\Lambda=3$ respectively, and the relations $R_{\mu\nu\alpha\beta}={R\over 12}\(g_{\mu\alpha}g_{\nu\beta}-g_{\mu\beta}g_{\nu\alpha}\)$ and $R_{\mu\nu}={R\over 4}g_{\mu\nu}$ valid for a maximally symmetric space.  
Given a suitable combination of these Lagrangians so that a constant field $\pi(x)=\pi_0=const.$ is a solution to the equations of motion, $f(\pi_0)= \rc \sin\(\pi_0\over \rc\)$ sets the radius of the de Sitter brane.  We call these Type I de Sitter DBI Galileons (see Figure \ref{types}).

\subsubsection{Killing vectors and symmetries}

Once again, we calculate the transformation laws using conformal inflationary coordinates $(u,y^i)$ on the de Sitter slices, even though they only cover half of each de Sitter slice.  The embedding becomes
\bea 
Y^0&=&\rc \sin\left(\rho\over \rc\right){1\over 2u}\left(1-u^2+y^2\right) \ ,\\
Y^1&=&\rc \cos\left(\rho\over \rc\right) \ ,\\
Y^2&=&\rc \sin\left(\rho\over \rc\right){1\over 2u}\left(1+u^2-y^2\right) \ ,\\
Y^{i+2}&=&\rc \sin\left(\rho\over \rc\right){y^i\over u},\ \ \ i=1,2,3 \ .
\eea
The coordinate ranges are $\rho\in (0,\pi\rc)$, $u\in(0,\infty)$ and $y^i\in(-\infty,\infty)$, and the induced metric then becomes
\be 
ds^2=d\rho^2+\rc^2  \sin^2\left(\rho\over \rc\right)\left[{1\over u^2}\left(-du^2+dy^2\right)\right] \ .
\ee

The 15 Lorentz generators of $M_6$
are all tangent to the $dS_5$ hyperboloid, and become the 15 isometries of its $so(5,1)$ isometry algebra.  Of these, 10 have no $\partial_\rho$ components and are parallel to the $dS_4$ foliation: these form the $so(4,1)$ isometry algebra of the $dS_4$ slices,
\bea 
-Y^2\partialb_0-Y^0\partialb_2&\rightarrow & d=u\partial_u+y^i\partial_i \ ,\\
-Y^{i+2}\partialb_0-Y^0\partialb_{i+2}&\rightarrow &j_i^+=uy_i\partial_u+\half\left(-1+u^2-y^2\right)\partial_i+y_iy^j\partial_j, \ \ \ i=1,2,3,\\
-Y^{i+2}\partialb_2+Y^2\partialb_{i+2}&\rightarrow& j_i^-=uy_i\partial_u + \half\left(1+u^2-y^2\right)\partial_i+y_iy^j\partial_j,  \ \ \ i=1,2,3,\\
Y^{i+2}\partialb_{j+2}-Y^{j+2}\partialb_{i+2}&\rightarrow &j_{ij}=y_i\partial_j-y_j\partial_i,\ \ \ \ i,j=1,2,3 . 
\eea
Taking the combinations
\bea 
p_i&=&j_i^+-j_i^-=-\partial_i \ ,\\
k_i&=& j_i^++j_i^-=2uy_i\partial_u+(u^2-y^2)\partial_i+2y_iy^j\partial_j \ ,
\eea
we then recognize $p_i$ and $j_{ij}$ as translations and rotations on the $y$ plane, while $d$ and $k_i$ fill out the $so(4,1)$ algebra.

The remaining 5 Killing vectors do have a $\partial_\rho$ component,
\bea 
-Y^1\partialb_0-Y^0\partialb_1&\rightarrow & K={\rc\over 2u}\left(1-u^2+y^2\right)\partial_\rho+\half\left(1+u^2+y^2\right)\cot\left(\rho\over \rc\right)\partial_u+u\cot\left(\rho\over \rc\right)y^i\partial_i \ ,\nn \\
-Y^2\partialb_1+Y^1\partialb_2&\rightarrow & K'={\rc\over 2u}\left(1+u^2-y^2\right)\partial_\rho+\half\left(1-u^2-y^2\right)\cot\left(\rho\over \rc\right)\partial_u-u\cot\left(\rho\over \rc\right)y^i\partial_i \ ,\nn \\
 -Y^{i+2}\partialb_1+Y^1\partialb_{i+2}&\rightarrow & K_i={\rc\over u}y_i\partial_\rho +y_i \cot\left(\rho\over \rc\right)\partial_u+u\cot\left(\rho\over \rc\right)\partial_i ,\ \ \ \ i=1,2,3. 
\eea

Defining the following linear combinations,
\bea 
K_+&=&K+K'={\rc\over u}\partial_\rho+ \cot\left(\rho\over \rc\right)\partial_u \ ,\nn \\
K_-&=&K-K'={\rc\over u}\left(-u^2+y^2\right)\partial_\rho+\left(u^2+y^2\right)\cot\left(\rho\over \rc\right)\partial_u+2u\cot\left(\rho\over \rc\right)y^i\partial_i \ ,\nn \\
K_i&=&{\rc\over u}y_i\partial_\rho +y_i \cot\left(\rho\over \rc\right)\partial_u+u\cot\left(\rho\over \rc\right)\partial_i \ ,
\eea
and using the relation $\delta_K\pi=K^5(x)-K^\mu(x,\pi)\partial_\mu\pi$ from~(\ref{specialcasesym}), we obtain the transformation rules
\bea
\delta_+\pi&=&{\rc\over u}- \cot\left(\pi\over \rc\right)\pi' \ ,\nn \\
\delta_-\pi&=&{\rc\over u}\left(-u^2+y^2\right)-\left(u^2+y^2\right)\cot\left(\pi\over \rc\right)\pi'-2u\cot\left(\pi\over \rc\right)y^i\partial_i\pi \ ,\nn \\
\delta_i\pi&=&{\rc\over u}y_i-y_i \cot\left(\pi\over \rc\right)\pi'-u\cot\left(\pi\over \rc\right)\partial_i\pi \ ,
\eea
where $\pi'\equiv\partial_u\pi$. The terms~(\ref{dSindSDBI}) are each invariant up to a total derivative under these transformations, and the symmetry breaking pattern is
\be 
so(5,1)\rightarrow so(4,1) \ .
\ee

%%%%%%%%%%%%%%%%%%%%
\subsection{A de Sitter brane in an anti-de Sitter bulk: $dS_4$ in $AdS_5$}

Using the description and notation for the $AdS_5$ embedding in section~\ref{MinAdSsection}, the following coordinates cover the intersection of the $AdS_5$ hyperbola  with the region $Y^0>{\cal R}$,
\bea 
\nn Y^0&=&{\cal R}\cosh \rho \ , \\ \nn
Y^1&=&{\cal R}\sinh \rho \sinh \tau \ , \\ \nn
Y^2&=&{\cal R}\sinh \rho \cosh \tau\cos\theta_1 \ , \\ \nn
Y^3&=&{\cal R}\sinh \rho \cosh \tau\sin\theta_1\cos\theta_2 \ , \\
Y^4&=&{\cal R}\sinh \rho \cosh \tau\sin\theta_1\sin\theta_2\cos\theta_3 \ , \\
Y^5&=&{\cal R}\sinh \rho \cosh \tau\sin\theta_1\sin\theta_2 \sin\theta_3 \ ,\\
\eea
where $\tau\in (-\infty,\infty)$, $\rho\in(0,\infty)$, and $\theta_i$, $i=1,2,3$ parametrize a $3$-sphere.  

The metric reads
\be
\label{adssynchronous2}
{ ds^2={\cal R}^2\left[d\rho^2+\sinh^2\rho\left(-d\tau^2+\cosh^2\tau\ d\Omega^2_{(3)}\right)\right]} \ .
\ee

Scaling $\rho$, the metric becomes $ds^2=d\rho^2+\rc^2\sinh^2\(\rho\over \rc\) ds^2_{dS_{4}}$, where $ds^2_{dS_{4}}$ is the global metric on a four-dimensional de Sitter space $dS_4$ of unit radius.  The foliation by $dS_{4}$ thus corresponds to $\rho={\rm constant}$ surfaces.  These slices are given by intersecting the planes $Y^0={\rm constant}$ with the hyperbola in the region $Y^0>{\cal R}$. (If we map $Y^0\rightarrow -Y^0$ then the coordinates cover the region $Y^0<-{\cal R}$, and the metric remains identical to~(\ref{adssynchronous2}), and this is the maximum extent to which we can extend the foliation.)

Comparing this with~(\ref{metricform}), we obtain  
\be
{f(\pi)=\rc\sinh (\pi/{\cal R}), \ \ \ g_{\mu\nu}=g_{\mu\nu}^{(dS_4)}} \ ,
\ee
and the terms~(\ref{generalterms}) become (without integration by parts)

\bea
{\cal L}_1&=& \sqrt{-g}{\rc^4\over 32}\( 12\ \pi-8\rc \sinh\(2\pi\over \rc\) +\rc \sinh\(4\pi\over \rc\)\) \ ,\\
{\cal L}_2&=&-\sqrt{-g}{\rc^4\over \gamma} \sinh^4\(\pi\over \rc\) \ ,\\
{\cal L}_3&=&\sqrt{-g}\left[\gamma^2 [\pi^3]-\rc^2[\Pi] \sinh^2\(\pi\over \rc\)+\rc^3(5-\gamma^2) \sinh^3\(\pi\over \rc\) \cosh\(\pi\over \rc\)\right] \ , \\
{\cal L}_4&=&\sqrt{-g}\Bigg[ {2\gamma^3\over \rc^2}\([\Pi][\pi^3]-[\pi^4]\)\csch^2\pir-{\gamma}\([\Pi]^2-[\Pi^2]+{8\gamma^2\over \rc}[\pi^3]\coth\pir\) \\
&&+\rc\gamma(4-\gamma^2)[\Pi]\sinh\pirt+{3\rc^2\over \gamma}\sinh^2\pir \(-2-3\gamma^2+\gamma^4+(2-3\gamma^2+\gamma^4)\cosh\pirt\)\Bigg] \ , \nn \\
{\cal L}_5&=&\sqrt{-g}\Bigg[ {3\gamma^4\over \rc^4}\(2([\pi^5]-[\Pi][\pi^4])+[\pi^3]([\Pi]^2-[\Pi^2])\)\csch^4\pir \nn\\ 
&&-{18\gamma^4\over \rc^3}\([\Pi][\pi^3]-[\pi^4]\)\csch^2\pir\coth\pir \nn\\
&&-{\gamma^2\over \rc^2}\csch^2\pir\([\Pi]^3-3[\Pi][\Pi^2]+2[\Pi^3]-{3\over 2}(3+10\gamma^2)[\pi^3]+{3\over 2}(3-10\gamma^2)[\pi^3]\cosh\pirt \) \nn \\
&&+{3\gamma^2\over \rc}(3-\gamma^2)([\Pi]^2-[\Pi^2])\coth\pir+{3\over 2}[\Pi] \(-3-10\gamma^2+4\gamma^4+(3-10\gamma^2+4\gamma^4)\cosh\pirt\) \nn \\
&& -{3\rc \over 4} \(-15-11\gamma^2+6\gamma^4+(15-17\gamma^2+6\gamma^4)\cosh\pirt\) \sinh\pirt \Bigg] \ , 
\label{dSinAdSDBI} 
\eea
where the background metric and covariant derivatives are those of the unit-radius $4$d de Sitter space, and 
\be 
\gamma={1\over \sqrt{1+{(\partial\pi)^2\over \rc^2 \sinh^2\(\pi\over \rc\)}}} \ .
\ee
Given suitable combinations of these Lagrangians so that a constant field $\pi(x)=\pi_0={\rm constant}$ is a solution to the equations of motion, $f(\pi_0)= \rc \sinh\(\pi_0\over \rc\)$ sets the radius of the de Sitter brane.  
We call these Type III de Sitter DBI Galileons (see Figure \ref{types}).

\subsubsection{Killing vectors and symmetries}

Once again we use conformal inflationary coordinates on the $dS_4$ slices.  The embedding becomes,
\bea 
Y^0&=&\rc \cosh\left(\rho\over \rc\right) \ ,\\
Y^1&=& \rc \sinh\left(\rho\over \rc\right){1\over 2u}\left(1-u^2+y^2\right) \ ,\\
Y^2&=&\rc \sinh\left(\rho\over \rc\right){1\over 2u}\left(1+u^2-y^2\right) \ ,\\
Y^{i+2}&=&\rc \sinh\left(\rho\over \rc\right){y^i\over u},\ \ \ i=1,2,3 \ ,
\eea
where $\rho\in (0,\infty)$ and $u\in(0,\infty)$.  The coordinate ranges are $\rho\in (0,\infty)$, $u\in(0,\infty)$, $y^i\in(-\infty,\infty)$, and the induced metric is
\be 
ds^2=d\rho^2+\rc^2  \sinh^2\left(\rho\over \rc\right)\left[{1\over u^2}\left(-du^2+dy^2\right)\right] \ .
\ee

The 15 Lorentz generators of $M_{4,2}$, $M_{AB}=Y_A\partialb_B-Y_B\partialb_A$, are all tangent to the $AdS_5$ hyperboloid, and become the 15 isometries of the $so(4,2)$ isometry algebra of $AdS_5$.  Of these, 10 have no $\partial_\rho$ components and are parallel to the $dS_4$ foliation.  These form the $so(4,1)$ isometry algebra of the $dS_4$ slices
\bea 
-Y^2\partialb_1-Y^1\partialb_2&\rightarrow & d=u\partial_u+y^i\partial_i \ ,\\
-Y^{i+2}\partialb_1-Y^1\partialb_{i+2}&\rightarrow &j_i^+=uy_i\partial_u+\half\left(-1+u^2-y^2\right)\partial_i+y_iy^j\partial_j, \ \ \ i=1,2,3, \\
-Y^{i+2}\partialb_2+Y^2\partialb_{i+2}&\rightarrow& j_i^-=uy_i\partial_u + \half\left(1+u^2-y^2\right)\partial_i+y_iy^j\partial_j,   \ \ \ i=1,2,3,\\
Y^{i+2}\partialb_{j+2}-Y^{j+2}\partialb_{i+2}&\rightarrow &j_{ij}=y_i\partial_j-y_j\partial_i , \ \ \ \ i,j=1,2,3.
\eea
Taking the combinations
\bea 
p_i&=&j_i^+-j_i^-=-\partial_i \ ,\\
k_i&=& j_i^++j_i^-=2uy_i\partial_u+(u^2-y^2)\partial_i+2y_iy^j\partial_j \ ,
\eea
we recognize $p_i$ and $j_{ij}$ as translations and rotations on the $y$ plane, with $d$ and $k_i$ filling out the rest of the $so(4,1)$ algebra.

The remaining 5 Killing vectors do have a $\partial_\rho$ component,
\bea 
Y^1\partialb_0-Y^0\partialb_1&\rightarrow & K={\rc\over 2u}\left(1-u^2+y^2\right)\partial_\rho+\half\left(1+u^2+y^2\right)\coth\left(\rho\over \rc\right)\partial_u+u\coth\left(\rho\over \rc\right)y^i\partial_i \ ,\nn \\
Y^2\partialb_0+Y^0\partialb_2&\rightarrow & K'={\rc\over 2u}\left(1+u^2-y^2\right)\partial_\rho+\half\left(1-u^2-y^2\right)\coth\left(\rho\over \rc\right)\partial_u-u\coth\left(\rho\over \rc\right)y^i\partial_i \ ,\nn \\
 Y^{i+2}\partialb_0+Y^0\partialb_{i+2}&\rightarrow & K_i={\rc\over u}y_i\partial_\rho +y_i \coth\left(\rho\over \rc\right)\partial_u+u\coth\left(\rho\over \rc\right)\partial_i, \ \ \ i=1,2,3. \nn
\eea

Taking the following linear combinations
\bea 
K_+&=&K+K'={\rc\over u}\partial_\rho+ \coth\left(\rho\over \rc\right)\partial_u \ ,\nn \\
K_-&=&K-K'={\rc\over u}\left(-u^2+y^2\right)\partial_\rho+\left(u^2+y^2\right)\coth\left(\rho\over \rc\right)\partial_u+2u\coth\left(\rho\over \rc\right)y^i\partial_i \ ,\nn \\
K_i&=&{\rc\over u}y_i\partial_\rho +y_i \coth\left(\rho\over \rc\right)\partial_u+u\coth\left(\rho\over \rc\right)\partial_i \ ,
\eea
and using the relation $\delta_K\pi=K^5(x)-K^\mu(x,\pi)\partial_\mu\pi$ from~(\ref{specialcasesym}), we obtain the transformation rules
\bea 
\delta_+\pi&=&{\rc\over u}- \coth\left(\pi\over \rc\right)\pi' \ ,\nn \\
\delta_-\pi&=&{\rc\over u}\left(-u^2+y^2\right)-\left(u^2+y^2\right)\coth\left(\pi\over \rc\right)\pi'-2u\coth\left(\pi\over \rc\right)y^i\partial_i\pi \ ,\nn \\
\delta_i\pi&=&{\rc\over u}y_i-y_i \coth\left(\pi\over \rc\right)\pi'-u\coth\left(\pi\over \rc\right)\partial_i\pi \ , \nn\\
\eea
where $\pi'\equiv\partial_u\pi$.  

The terms~(\ref{dSinAdSDBI}) are each invariant up to a total derivative under these transformations, and the symmetry breaking pattern is
\be 
so(4,2)\rightarrow so(4,1) \ .
\ee

%%%%%%%%%%%%%%%%%%%%
\subsection{An anti-de Sitter brane in an anti-de Sitter bulk: $AdS_4$ in $AdS_5$}

Using the description and notation for the $AdS_5$ embedding from section~\ref{MinAdSsection}, hyperbolic coordinates on $AdS^5$ are
\bea 
\nn Y^0&=&{\cal R}\cos \tau \cosh \rho\cosh\psi \ , \\ \nn
Y^1&=&{\cal R}\sin \tau \cosh \rho\cosh\psi \ , \\ \nn
Y^2&=&{\cal R}\sinh\rho \ , \\ \nn
Y^3&=&{\cal R}\cosh \rho\sinh\psi\cos\theta_1 \ , \\ \nn
Y^4&=&{\cal R}\cosh \rho\sinh\psi\sin\theta_1\cos\theta_2 \ , \\
Y^5&=&{\cal R}\cosh \rho\sinh\psi\sin\theta_1\sin\theta_2 \ ,
\eea
where $\tau\in(-\pi,\pi)$ (the universal cover is obtained by extending this to $\tau\in(-\infty,\infty)$), $\rho\in(-\infty,\infty)$, $\psi\in(0,\infty)$, and $\theta_1,\theta_2$ parametrize a $2$-sphere.  These coordinates cover the entire $AdS_5$ hyperbola, and after extending $\tau$, the whole of $AdS_5$.  

The metric reads
\be
\label{adshyperbolic}
{ ds^2={\cal R}^2\left[d\rho^2+\cosh^2\rho\left(-\cosh^2\psi\ d\tau^2+d\psi^2+\sinh^2\psi\ d\Omega^2_{(2)}\right)\right]} \ ,
\ee
and after scaling $\rho$, this becomes $ds^2=d\rho^2+\rc^2\cosh^2\(\rho\over \rc\) ds^2_{AdS_{4}},$ where $ds^2_{AdS_{4}}$ is the global metric on an anti-de Sitter space $AdS_4$ of unit radius.  The foliation by $AdS_{4}$ thus corresponds to $\rho={\rm constant}$ surfaces, and these slices are given by intersecting the planes $Y^2={\rm constant}$ with the hyperbola.  This foliation covers the entire $AdS_5$ space.

Comparing this with~(\ref{metricform}), we obtain  
\be
{f(\pi)=\rc\cosh (\pi/{\cal R}), \ \ \ g_{\mu\nu}=g_{\mu\nu}^{(AdS_4)}} \ ,
\ee
and the terms~(\ref{generalterms}) become (without any integrations by parts)

\bea 
{\cal L}_1&=&\sqrt{-g}{\rc^4\over 32}\( 12\ \pi+8\rc \sinh\(2\pi\over \rc\) +\rc \sinh\(4\pi\over \rc\)\) \ ,\\
{\cal L}_2&=&-\sqrt{-g}{\rc^4\over \gamma} \cosh^4\(\pi\over \rc\) \ ,\\
{\cal L}_3&=&\sqrt{-g}\left[\gamma^2 [\pi^3]-\rc^2[\Pi] \cosh^2\(\pi\over \rc\)+\rc^3(5-\gamma^2) \cosh^3\(\pi\over \rc\) \sinh\(\pi\over \rc\)\right] \ , \\
{\cal L}_4&=&\sqrt{-g}\Bigg[ {2\gamma^3\over \rc^2}\([\Pi][\pi^3]-[\pi^4]\)\sech^2\pir-{\gamma}\([\Pi]^2-[\Pi^2]+{8\gamma^2\over \rc}[\pi^3]\tanh\pir\) \\
&&+\rc\gamma(4-\gamma^2)[\Pi]\sinh\pirt+{3\rc^2\over \gamma}\cosh^2\pir \(2+3\gamma^2-\gamma^4+(2-3\gamma^2+\gamma^4)\cosh\pirt\)\Bigg] \ , \nn \\
{\cal L}_5&=&\sqrt{-g}\Bigg[ {3\gamma^4\over \rc^4}\(2([\pi^5]-[\Pi][\pi^4])+[\pi^3]([\Pi]^2-[\Pi^2])\)\sech^4\pir \nn\\ 
&&-{18\gamma^4\over \rc^3}\([\Pi][\pi^3]-[\pi^4]\)\sech^2\pir\tanh\pir \nn\\
&&-{\gamma^2\over \rc^2}\sech^2\pir\([\Pi]^3-3[\Pi][\Pi^2]+2[\Pi^3]+{3\over 2}(3+10\gamma^2)[\pi^3]+{3\over 2}(3-10\gamma^2)[\pi^3]\cosh\pirt \) \nn \\
&&+{3\gamma^2\over \rc}(3-\gamma^2)([\Pi]^2-[\Pi^2])\tanh\pir+{3\over 2}[\Pi] \(3+10\gamma^2-4\gamma^4+(3-10\gamma^2+4\gamma^4)\cosh\pirt\) \nn \\
&& -{3\rc \over 4} \(15+11\gamma^2-6\gamma^4+(15-17\gamma^2+6\gamma^4)\cosh\pirt\) \sinh\pirt \Bigg] \ , 
\label{AdSinAdSDBI}
\eea
where the background metric and covariant derivatives are those of a unit-radius $AdS_4$, and  
\be 
\gamma={1\over \sqrt{1+{(\partial\pi)^2\over \rc^2 \cosh^2\(\pi\over \rc\)}}} \ .
\ee
In evaluating~(\ref{dSinMDBI}), we have used that fact that the scalar curvature and cosmological constant of the unit-radius $AdS_4$ are
$R=-12$ and $\Lambda=-3$ respectively, as well as the relations $R_{\mu\nu\alpha\beta}={R\over 12}\(g_{\mu\alpha}g_{\nu\beta}-g_{\mu\beta}g_{\nu\alpha}\)$, $R_{\mu\nu}={R\over 4}g_{\mu\nu}$ valid for a maximally symmetric space.  Given suitable combinations of these Lagrangians so that a constant field $\pi(x)=\pi_0={\rm constant}$ is a solution to the equations of motion, $f(\pi_0)= \rc \cosh\(\pi_0\over \rc\)$ sets the radius of the anti-de Sitter brane.  We call these anti-de Sitter DBI Galileons (see Figure \ref{types}).

\subsubsection{Killing vectors and symmetries}

We use Poincare coordinates $(u,x^0,x^1,x^2)$ on the $AdS_4$ slices.  The embedding becomes,
\bea 
Y^0&=&\rc \cosh\left(\rho\over \rc\right){1\over 2u}\left(1+u^2+x^2\right) \ ,\\
Y^1&=& \rc \cosh\left(\rho\over \rc\right){x^0\over u} \ ,\\
Y^2&=&\rc \sinh\left(\rho\over \rc\right) \ ,\\
Y^3&=& \rc \cosh\left(\rho\over \rc\right){1\over 2u}\left(1-u^2-x^2\right) \ ,\\
Y^{i+3}&=&\rc \cosh\left(\rho\over \rc\right){x^i\over u},\ \ \ i=1,2 \ .
\eea
Here $x^2\equiv \eta_{ij}x^ix^j$, where $\eta_{ij}={\rm diag}(-1,1,1)$ is the Minkowski 3-metric.  The coordinate ranges are $\rho\in (0,\infty)$, $u\in(0,\infty)$ and $x^i\in(-\infty,\infty)$, and the induced metric is
\be 
ds^2=d\rho^2+\rc^2  \cosh^2\left(\rho\over \rc\right)\left[{1\over u^2}\left(du^2+\eta_{ij}dx^idx^j\right)\right] \ .
\ee

The 15 Lorentz generators of $M_{4,2}$
are all tangent to the $AdS_5$ hyperboloid, and become the 15 isometries of the $so(4,2)$ isometry algebra of $AdS_5$.  Of these, 10 have no $\partial_\rho$ components and are parallel to the $AdS_4$ foliation - these form the $so(3,2)$ isometry algebra of the $AdS_4$ slices,
\bea 
-Y^3\partialb_0-Y^0\partialb_3&\rightarrow & u\partial_u+x^i\partial_i \ ,\nn \\
-Y^1\partialb_0+Y^0\partialb_1&\rightarrow & u x^0\partial_u+\half\left(1+u^2+x^2\right)\partial_0+x^0x^j\partial_j \ ,\nn \\
-Y^{i+3}\partialb_0-Y^0\partialb_{i+3}&\rightarrow & u x_i\partial_u-\half\left(1+u^2+x^2\right)\partial_i+x_ix^j\partial_j,\ \ \ i=1,2\nn \\
-Y^3\partialb_1-Y^1\partialb_3&\rightarrow & u x^0\partial_u+\half\left(-1+u^2+x^2\right)\partial_0+x^0x^j\partial_j \ ,\nn \\
-Y^{i+3}\partialb_3+Y^3\partialb_{i+3}&\rightarrow & u x_i\partial_u-\half\left(-1+u^2+x^2\right)\partial_i+x_ix^j\partial_j,\ \ \ i=1,2 \nn \\
Y^{i+3}\partialb_1+Y^1\partialb_{i+3}&\rightarrow &x^i\partial_0+x^0\partial_i,\ \ \ \ i=1,2 \nn \\
Y^{5}\partialb_4+Y^4\partialb_{5}&\rightarrow & x^2\partial_1-x^1\partial_2 \ ,
\eea
where the sums are over $j=0,1,2$, and indices are raised and lowered with $\eta_{ij}$.  These may be grouped as
\bea 
d&=&u\partial_u+x^i\partial_i \ ,\\
j_i^+&=&ux_i\partial_u-\half\left(1+u^2+x^2\right)\partial_i+x_ix^j\partial_j,\ \ \ \ i=0,1,2\\
 j_i^-&=&ux_i\partial_u - \half\left(-1+u^2+x^2\right)\partial_i+x_ix^j\partial_j,\ \ \ \ i=0,1,2\\
j_{ij}&=&x_i\partial_j-x_j\partial_i, \ \ \ \ i,j=0,1,2 \ ,
\eea
and by taking the combinations
\bea 
p_i&=&j_i^+-j_i^-=-\partial_i \ ,\\
k_i&=& j_i^++j_i^-=2ux_i\partial_u-(u^2+x^2)\partial_i+2x_ix^j\partial_j \ ,
\eea
we recognize $p_i$ and $j_{ij}$ as translations and rotations on the $x$-space, with $d$ and $k_i$ filling out the rest of the $so(3,2)$ algebra.

The remaining 5 Killing vectors do have a $\partial_\rho$ component,
\bea 
Y^2\partialb_0+Y^0\partialb_2&\rightarrow & K={\rc\over 2u}\left(1+u^2+x^2\right)\partial_\rho+\half\left(1-u^2+x^2\right)\tanh\left(\rho\over \rc\right)\partial_u-u\tanh\left(\rho\over \rc\right)x^i\partial_i \ ,\nn \\
Y^3\partialb_2-Y^2\partialb_3&\rightarrow & K'={\rc\over 2u}\left(1-u^2-x^2\right)\partial_\rho+\half\left(1+u^2-x^2\right)\tanh\left(\rho\over \rc\right)\partial_u+u\tanh\left(\rho\over \rc\right)x^i\partial_i \ ,\nn \\
Y^2\partialb_1+Y^1\partialb_2&\rightarrow & {\rc\over u}x^0\partial_\rho+x^0\tanh\left(\rho\over \rc\right)\partial_u+u\tanh\left(\rho\over \rc\right)\partial_0 \ ,\nn \\
 Y^{i+3}\partialb_2-Y^2\partialb_{i+3}&\rightarrow & {\rc\over u}x^i\partial_\rho+x^i\tanh\left(\rho\over \rc\right)\partial_u-u\tanh\left(\rho\over \rc\right)\partial_i,\ \ \ i=1,2 \ ,\nn
\eea
which may be combined to form
\bea 
K_+&=&K+K'={\rc\over u}\partial_\rho+ \tanh\left(\rho\over \rc\right)\partial_u \ ,\nn \\
K_-&=&K-K'={\rc\over u}\left(u^2+x^2\right)\partial_\rho+\left(-u^2+x^2\right)\tanh\left(\rho\over \rc\right)\partial_u-2u\tanh\left(\rho\over \rc\right)x^i\partial_i \ ,\nn \\
K_i&=&{\rc\over u}x_i\partial_\rho +x_i \tanh\left(\rho\over \rc\right)\partial_u-u\tanh\left(\rho\over \rc\right)\partial_i,\ \ \ i=0,1,2.
\eea
Using the relation $\delta_K\pi=K^5(x)-K^\mu(x,\pi)\partial_\mu\pi$ from~(\ref{specialcasesym}), we obtain the transformation rules
\bea 
\delta_+\pi&=&{\rc\over u}- \tanh\left(\pi\over \rc\right)\pi' \ , \\
\delta_-\pi&=&{\rc\over u}\left(u^2+x^2\right)-\left(-u^2+x^2\right)\tanh\left(\pi\over \rc\right)\pi'+2u\tanh\left(\pi\over \rc\right)x^i\partial_i\pi \ , \\
\delta_i\pi&=&{\rc\over u}x_i-x_i \tanh\left(\pi\over \rc\right)\pi'+u\tanh\left(\pi\over \rc\right)\partial_i\pi,\ \ \ i=0,1,2 \ ,
\label{AdSinAdStrans}
\eea
where $\pi'\equiv\partial_u\pi$. 

The terms~(\ref{AdSinAdSDBI}) are each invariant up to a total derivative under these transformations, and the symmetry breaking pattern is
\be 
so(4,2)\rightarrow so(3,2) \ .
\ee

\section{Small field limits: the analogues of Galileons}
\label{sec:smallfieldlimits}

The Lagrangians we have uncovered have a fairly complicated, non-polynomial form.  We know in the Minkowski case that
the special case of the Galileon symmetry arises in a particular limit \cite{deRham:2010eu}, and that this limit greatly simplifies the actions.  In this section, we consider similar limits for the general theories we have constructed.

Consider a Lagrangian ${\cal L}$ that may be expanded in some formal series in a parameter $\lambda$ as
\be 
{\cal L}=\lambda^n\left({\cal L}_{(0)}+\lambda {\cal L}_{(1)}+\lambda^2{\cal L}_{(2)}+\cdots\right) \ ,
\ee
where $n$ is an integer, indicating that the series need not start at order $\lambda^0$.  Suppose ${\cal L}$ possesses a symmetry that may also be expanded in such a series
\be 
\delta\pi=\lambda^m\left(\delta_{(0)}\pi+\lambda \delta_{(1)}\pi+\lambda^2 \delta_{(2)}\pi+\cdots\right) \ ,
\ee
where $m$ is another integer, again indicating that this series also need not start at order $\lambda^0$.  The statement that $\delta\pi$ is a symmetry of ${\cal L}$ is
\be 
\label{invarianceeq} {\delta^{EL}{\cal L}\over \delta \pi}\delta\pi \simeq 0 \ ,
\ee
where ${\delta^{EL}{\cal L}\over \delta \pi}$ is the Euler-Lagrange derivative and $\simeq$ indicates equality up to a total derivative.  

Expanding~(\ref{invarianceeq}) in powers of $\lambda$ yields a series of equations
\bea 
&& {\delta^{EL}{\cal L}_{(0)}\over \delta \pi}\delta_{(0)} \pi\simeq 0 \ , \\
&& {\delta^{EL}{\cal L}_{(1)}\over \delta \pi}\delta_{(0)}\pi+ {\delta^{EL}{\cal L}_{(0)}\over \delta \pi}\delta_{(1)}\pi \simeq 0 \ , \nn\\
&&\vdots
\eea
with the first of these indicating that $\delta_{(0)}$ is a symmetry of ${\cal L}_{(0)}$. Our goal in this section is to seek expansions of this form for the various examples we have constructed, in order to find simpler, but still non-trivial, theories with the same number of symmetries.  

The expansion we choose is one in powers of the field $\pi$ around some background.  We expand $\pi$ around a constant background value $\pi_0$ and let $\lambda$ count powers of the deviation from this background; i.e. we make the replacement
\be 
\pi\rightarrow \pi_0+\lambda\pi \ ,
\ee
and then expand the Lagrangians and symmetries in powers of $\lambda$.

Applying this small field limit to the DBI Galileons~(\ref{DBIGalileonterms}) gives rise to the original Galileons first studied in~\cite{Nicolis:2008in}.  These are, up to total derivatives,
\begin{align}
\mathcal{L}_{2}&=\pi \ , \nn \\
\mathcal{L}_{2}&=-{1\over 2}(\partial\pi)^2 \ , \nn \\
\mathcal{L}_{3}&=-{1\over 2}(\partial\pi)^2[\Pi] \ ,\nn  \\
\mathcal{L}_{4}& =-{1\over 2}(\partial\pi)^2\left([\Pi]^2-[\Pi^2]\right) \ ,\nn \\
\mathcal{L}_{5}& =-{1\over 2}(\partial\pi)^2\left([\Pi]^3-3[\Pi][\Pi^2]+2[\Pi^3]\right)\ .
\label{normalGalileons}
\end{align}
Note that lower order terms in the expansion are total derivatives.  For example, in the expansion of ${\cal L}_4$ there exists an ${\cal O}\(\pi^2\)$ piece, but this is a total derivative in Minkowski space, and the first non-trivial term is the ${\cal O}\(\pi^4\)$ piece shown above. 

Applying the small field limit to the transformation laws~(\ref{DBIGalileontrans}) yields 
\bea 
&&\delta\pi=1 \ , \nn \\
&& \delta_\mu \pi=x_\mu \ ,
\label{normalGalileontrans}
\eea
under which the terms~(\ref{normalGalileons}) are invariant.  This is the original Galilean symmetry considered in \cite{Nicolis:2008in}.  The small field limit can also be applied to the case of a flat brane embedded in an $AdS_5$ bulk~(\ref{conformalDBIGalileonterms}), but the resulting actions and transformation laws are identical to those of~(\ref{normalGalileons}),~(\ref{normalGalileontrans}).

Applying this technique to a de Sitter brane embedded in a flat bulk, we expand~(\ref{dSinMDBI}) around some constant background. The following linear combinations allow us to successively cancel the lowest order terms in $\lambda$ up to total derivatives on $dS_4$, yielding terms which start at order $\lambda$, $\lambda^2$, etc.
\bea  
\bar{\cal L}_1&=&{1\over \pi_0^4}{\cal L}_1=\sqrt{-g}\pi \ , \nn\\
\bar{\cal L}_2&=&{1\over \pi_0^2}\left( {\cal L}_2+{4\over\pi_0}{\cal L}_1\right)=-\half\sqrt{-g} \left((\partial\pi)^2-4 \pi^2\right) \ ,\nn \\
\bar{\cal L}_3&=& {\cal L}_3+{6\over\pi_0}{\cal L}_2+{12\over\pi_0^2}{\cal L}_1=\sqrt{-g}\left(-{1\over 2}(\partial\pi)^2[\Pi]-3 (\partial\pi)^2\pi+4\pi^3\right) \ ,\nn\\
\bar{\cal L}_4&=& \pi_0^2\left( {\cal L}_4+{6\over\pi_0}{\cal L}_3+{18\over\pi_0^2}{\cal L}_2+{24\over\pi_0^3}{\cal L}_1\right) \nn \\ 
&=&\sqrt{-g}\left[-\half(\partial\pi)^2\left([\Pi]^2-[\Pi^2]+\half(\partial\pi)^2+6\pi[\Pi]+18\pi^2\right)+6\pi^4\right] \ , \nn \\
\bar{\cal L}_5&=&  \pi_0^4\left( {\cal L}_5+{4\over\pi_0}{\cal L}_4+{12\over\pi_0^2}{\cal L}_3+{24\over\pi_0^3}{\cal L}_2+{24\over\pi_0^4}{\cal L}_1\right) \nn \\ 
&=& \sqrt{-g}\left[-\half\left((\partial\pi)^2+{1\over 5}\pi^2\right)\left([\Pi]^3-3[\Pi][\Pi^2]+2[\Pi^3]\right)\right. \nn \\ 
&&\left.-{12\over 5}\pi(\partial\pi)^2\left([\Pi]^2-[\Pi^2]+{27\over 12}[\Pi]\pi+5\pi^2\right)+{24\over 5}\pi^5\right] \ .
\eea
Scaling the coordinates to $(\hat u,\hat y^i)\equiv(L u, Ly^i)$, carrying dimensions of length, the $dS_4$ curvature becomes $R={12\over L^2}$, and canonically normalizing the field to $\hat\pi={1\over L^2}\pi$, we then obtain
\begin{eqnarray} 
\hat{\cal L}_1&=&\sqrt{-g}\hat\pi \ , \nn\\
\hat{\cal L}_2&=&-\half\sqrt{-g} \left((\partial\hat\pi)^2-{4\over L^2}\hat \pi^2\right) \ ,\nn \\
\hat{\cal L}_3&=& \sqrt{-g}\left(-{1\over 2}(\partial\hat\pi)^2[\hat\Pi]-{3\over L^2} (\partial\hat\pi)^2\hat\pi+{4\over L^4}\hat\pi^3\right) \ ,\nn\\
\hat{\cal L}_4&=&\sqrt{-g}\left[-\half(\partial\hat\pi)^2\left([\hat\Pi]^2-[\hat\Pi^2]+{1\over 2L^2}(\partial\hat\pi)^2+{6\over L^2}\hat\pi[\hat\Pi]+{18\over L^4}\hat\pi^2\right)+{6\over L^6}\hat\pi^4\right] \ , \nn \\
\hat{\cal L}_5&=& \sqrt{-g}\left[-\half\left((\partial\hat\pi)^2+{1\over 5L^2} \hat\pi^2\right)\left([\hat\Pi]^3-3[\hat\Pi][\hat\Pi^2]+2[\hat\Pi^3]\right)\right. \nn \\ 
&&\left.-{12\over 5L^2} \hat\pi(\partial \hat\pi)^2\left([\hat\Pi]^2-[\hat\Pi^2]+{27\over 12L^2}[\hat\Pi] \hat\pi+{5\over L^4} \hat\pi^2\right)+{24\over 5L^8} \hat\pi^5\right]  \ ,
\label{dsGalileonsscaled}
\end{eqnarray}
where $\hat{\cal L}_n={1\over L^{4n+2}}\bar{\cal L}_n$.

These expressions are invariant under the lowest order symmetry transformations obtained by taking the small field limit of~(\ref{dSinMtrans}),
\bea 
\delta_{+}\hat\pi&=&{1\over u}\left(u^2-y^2\right) \ ,\nn\\
\delta_{-} \hat\pi&=&-{1\over u} \ ,\nn\\ 
\label{dSGalileontrans}
\delta_{i} \hat\pi &=& {y_i\over u} \ .
\eea
The terms~(\ref{dsGalileonsscaled}) are Galileons which naturally live in de Sitter space, and become the original Galileons in the limit where the $dS_4$ radius goes to infinity.  They have the same number of nonlinear shift-like symmetries as the original flat space Galileons, despite the fact that they live on a curved space.  As such, we anticipate them being naturally suited to models of inflation and dark energy.

Another fascinating new feature that is not shared by the original Galileons is the existence of a potential.  In particular, the quadratic term $\hat{\cal L}_2$ comes with a mass term of order the $4$d de Sitter radius.  The symmetries (\ref{dSGalileontrans}) fix the value of the mass (in fact, each of the symmetries in (\ref{dSGalileontrans}) is alone sufficient to fix the mass).  If the coefficient of $\hat{\cal L}_2$ is chosen to be positive, so that the scalar field is not a ghost, then this mass is tachyonic. However, this instability is not necessarily worrisome because its timescale is of order the de Sitter time.  Furthermore, this small mass should not be renormalized, because its value is protected by symmetry.  The higher terms also come with cubic, quartic, and quintic terms in the potential, with values tied to the kinetic structure by the symmetries.

The small field limit may also be applied to the examples of a de Sitter brane embedded in either a de Sitter~(\ref{dSindSDBI}) or anti-de Sitter~(\ref{dSinAdSDBI}) bulk.  The resulting actions and transformation laws are identical to those of~(\ref{dsGalileonsscaled}) and~(\ref{dSGalileontrans}).

Finally, we apply the small field expansion to the case of an anti-de Sitter brane embedded in an anti-de Sitter bulk, by expanding the terms~(\ref{AdSinAdSDBI}) around a constant background $\pi_0$.  In a similar manner to the previous case, the following linear combinations yield terms which start at order $\lambda$, $\lambda^2$, etc. up to total derivatives.
\bea  
\bar{\cal L}_1&=&{1\over L^4}{\cal L}_1=\sqrt{-g}\pi \ ,\nn \\
\bar{\cal L}_2&=&{1\over L^2}\left[ {\cal L}_2+{4\over\rc}\tanh\left(\pi_0\over \rc\right){\cal L}_1\right]=-\half\sqrt{-g}\left((\partial\pi)^2+4 \pi^2\right) \ ,\nn \\
\bar{\cal L}_3&=& {\cal L}_3+{6\over\rc}\tanh\left(\pi_0\over \rc\right){\cal L}_2+{4\over \rc^2}\left(2-3\ \text{sech}^2\left(\frac{\pi_0}{\rc}\right)\right){\cal L}_1=\sqrt{-g}\left(-{1\over 2}(\partial\pi)^2[\Pi]+3 (\partial\pi)^2\pi+4\pi^3\right) \ ,\nn \\
\bar{\cal L}_4&=&L^2\left[ {\cal L}_4+ {6\over\rc}\tanh\left(\pi_0\over \rc\right){\cal L}_3+ {6\over \rc^2}\left(4-3\ \text{sech}^2\left(\frac{\pi_0}{\rc}\right)\right){\cal L}_2-{24\over \rc^3} \text{sech}^2\left(\frac{\pi_0}{\rc}\right)\tanh\left(\pi_0\over \rc\right){\cal L}_1\right]\nn \\ 
&=& \sqrt{-g}\left[-\half(\partial\pi)^2\left([\Pi]^2-[\Pi^2]-\half(\partial\pi)^2-6\pi[\Pi]+18\pi^2\right)-6\pi^4\right] \ ,\nn \\
\bar{\cal L}_5&=& L^4\left[ {\cal L}_5+ {4\over\rc}\tanh\left(\pi_0\over \rc\right){\cal L}_4+ {3\over \rc^2}\left(5-4\ \text{sech}^2\left(\frac{\pi_0}{\rc}\right)\right){\cal L}_3\right. \nn\\ 
&& \left.+{12\over \rc^3} \text{sech}^3\left(\frac{\pi_0}{\rc}\right)\left(\sinh\left(3\pi_0\over \rc\right)-\sinh\left(\pi_0\over \rc\right)\right){\cal L}_2+{24\over \rc^4} \text{sech}^4\left(\frac{\pi_0}{\rc}\right){\cal L}_1\right]\nn \\ 
&=& \sqrt{-g}\left[-\half\left((\partial\pi)^2-{1\over 5}\pi^2\right)\left([\Pi]^3-3[\Pi][\Pi^2]+2[\Pi^3]\right)\right. \nn \\ 
&&\left.+{12\over 5}\pi(\partial\pi)^2\left([\Pi]^2-[\Pi^2]-{27\over 12}[\Pi]\pi+5\pi^2\right)+{24\over 5}\pi^5\right] \ ,
\eea
where $L=\rc\cosh^4\left(\pi_0\over \rc\right)$ is the $AdS_{3,1}$ radius. 

Scaling the coordinates to $(\hat u,\hat x^i)\equiv(L u, Ly^i)$ so that they carry dimensions of length, the $AdS_4$ curvature becomes $R=-{12\over L^2}$, and canonically normalizing the field to $\hat\pi={1\over L^2}\pi$, we then obtain
\begin{eqnarray} 
\hat{\cal L}_1&=&\sqrt{-g}\hat\pi \ , \nn\\
\hat{\cal L}_2&=&-\half\sqrt{-g} \left((\partial\hat\pi)^2+{4\over L^2}\hat \pi^2\right) \ ,\nn \\
\hat{\cal L}_3&=& \sqrt{-g}\left(-{1\over 2}(\partial\hat\pi)^2[\hat\Pi]+{3\over L^2} (\partial\hat\pi)^2\hat\pi+{4\over L^4}\hat\pi^3\right) \ ,\nn\\
\hat{\cal L}_4&=&\sqrt{-g}\left[-\half(\partial\hat\pi)^2\left([\hat\Pi]^2-[\hat\Pi^2]-{1\over 2L^2}(\partial\hat\pi)^2-{6\over L^2}\hat\pi[\hat\Pi]+{18\over L^4}\hat\pi^2\right)-{6\over L^6}\hat\pi^4\right] \ , \nn \\
\hat{\cal L}_5&=& \sqrt{-g}\left[-\half\left((\partial\hat\pi)^2-{1\over 5L^2} \hat\pi^2\right)\left([\hat\Pi]^3-3[\hat\Pi][\hat\Pi^2]+2[\hat\Pi^3]\right)\right. \nn \\ 
&&\left.+{12\over 5L^2} \hat\pi(\partial \hat\pi)^2\left([\hat\Pi]^2-[\hat\Pi^2]-{27\over 12L^2}[\hat\Pi] \hat\pi+{5\over L^4} \hat\pi^2\right)+{24\over 5L^8} \hat\pi^5\right] \ ,
\label{AdSGalileonsscaled}
\end{eqnarray}
where $\hat{\cal L}_n={1\over L^{4n+2}}\bar{\cal L}_n$.

These terms are invariant under the lowest order symmetry transformations obtained by taking the small field limit of~(\ref{AdSinAdStrans})
\bea 
\delta_{+(0)}\hat\pi&=&{\rc\over u} \ ,\nn \\
\delta_{-(0)}\hat\pi&=&{\rc\over u}\left(u^2+x^2\right) \ ,\nn \\
\delta_{i(0)}\hat\pi&=&{\rc\over u}x_i,\ \ \ i=0,1,2  \ .\nn\\
\eea
These are Galileons that live on anti-de Sitter space.  In this case, the quadratic term comes with a non-tachyonic mass of order the $AdS_4$ radius.  

While we have focused on the construction of new effective field theories through the small field expansion of embedded brane models, it is important to note that there may well exist other expansions that lead to different theories in the limit.  For the example of a flat brane embedded in an anti-de Sitter bulk (\ref{conformalDBIGalileonterms}), the theory admits an expansion in powers of derivatives.  Up to total derivatives, the derivative expansion yields
\bea  
\bar{\cal L}_1&=&{1\over \rc}{\cal L}_1=-{1\over 4}e^{-4\hat\pi} \ , \nn\\
\bar{\cal L}_2&=&{1\over \rc^2}\left( {\cal L}_2-{4\over \rc}{\cal L}_1\right)=-\half e^{-2\hat\pi}(\partial\hat\pi)^2 \ ,\nn \\
\bar{\cal L}_3&=&{1\over \rc^3}\left(  {\cal L}_3-{6\over \rc}{\cal L}_2+{8\over \rc^2}{\cal L}_1\right)=-\frac{1}{2}(\partial \hat{\pi})^2 \Box \hat{\pi}+\frac{1}{4} (\partial \hat{\pi})^4 \ ,\nn\\
\bar{\cal L}_4&=& {1\over \rc^4}\left( {\cal L}_4-{6\over \rc}{\cal L}_3+{24\over \rc^2}{\cal L}_2\right) \nn \\ 
&=&- \frac{1}{2}e^{2\hat \pi}(\partial \hat \pi)^2
\([\hat \Pi]^2-[\hat \Pi^2]+{2\over 5}((\partial \hat \pi)^2 \Box \hat \pi-[\hat \pi^3])+{3\over 10}(\partial \hat \pi)^4\) \ , \nn \\
\bar{\cal L}_5&=&  {1\over \rc^5}\left( {\cal L}_5-{4\over \rc}{\cal L}_4+{15\over \rc^2}{\cal L}_3-{48\over \rc^3}{\cal L}_2\right) \nn \\ 
&=& -\half e^{4\hat \pi} (\partial \hat \pi)^2
\Big[[\hat \Pi]^3-3[\hat \Pi][\hat \Pi^2]+2[\hat \Pi^3]+3(\partial \hat \pi)^2([\hat \Pi]^2-[\hat \Pi^2])\nn\\
&&+\frac{30}{7}(\partial \hat \pi)^2((\partial \hat \pi)^2[\hat \Pi]-[\hat \pi^3])-\frac{3}{28}(\partial \hat \pi)^6\Big] \ ,\label{DBIgalderivexpan}
\eea
where $\hat\pi\equiv\pi/\rc$. These are the conformal Galileons~\cite{Nicolis:2008in,deRham:2010eu,Khoury:2011da}.  Their transformation laws come from applying the derivative expansion to the transformation laws (\ref{MinAdStrans}),
 \bea 
\delta\hat\pi&=&1-x^\mu\partial_\mu\hat\pi,\nn\\
\delta_\mu\hat\pi&=&2x_\mu+x^2\partial_\mu\hat\pi-2x_\mu x^\nu\partial_\nu\hat\pi \ . \label{MinAdStranslim}
\eea
In taking the limit in powers of derivatives, we must remember that the explicit factors of the coordinates in the transformation laws are assigned a power of inverse derivatives.  The terms (\ref{DBIgalderivexpan}) are each invariant up to a total derivative under (\ref{MinAdStranslim}).  As mentioned in \cite{deRham:2010eu}, it is remarkable that this limit does not alter the commutation relations of the symmetries, so that the algebra remains $so(4,2)$.

The derivative expansion can also be applied to the DBI Galileons (\ref{DBIGalileonterms}).  The result is identical to the small field limit, since the powers of $\pi$ and powers of $\partial$ within each limiting Lagrangian are identical.  

A derivative expansion does not, however, seem applicable in general.  To see the problem, attempt to construct an order four derivative term from the general Lagrangians in (\ref{generalterms}).  It is necessary to find a constant $A$ such that the two derivative part in the expression ${\cal L}_3+A{\cal L}_2$ is a total derivative.  The two derivative part reads $\sqrt{-g}\(3ff'-{A\over 2}f^2\)(\partial\pi)^2$, up to a total derivative, and for this to vanish we must have $f\propto e^{A\pi/6}$.  The only cases of ours that conform to this are the conformal DBI Galileons ($A\not= 0$) and the ordinary DBI Galileons ($A=0$).

\subsection{Symmetry breaking and ghosts}

By writing the actions of the previous section in terms of the scalar curvature, $R={12\over L^2}$ for $dS_4$, $R=-{12\over L^2}$ for $AdS_4$, and $R=0$ for $M_4$, it is possible to combine the $dS_4$ Galileons (\ref{dsGalileonsscaled}), the $AdS_4$ Galileons (\ref{AdSGalileonsscaled}) and the flat space Galileons (\ref{normalGalileons}) into the single set of expressions

\begin{eqnarray} 
\hat{\cal L}_1&=&\sqrt{-g}\hat\pi \ , \nn\\
\hat{\cal L}_2&=&-\half\sqrt{-g} \left((\partial\hat\pi)^2-{R\over 3}\hat \pi^2\right) \ ,\nn \\
\hat{\cal L}_3&=& \sqrt{-g}\left(-{1\over 2}(\partial\hat\pi)^2[\hat\Pi]-{R\over 4} (\partial\hat\pi)^2\hat\pi+{R^2\over 36}\hat\pi^3\right) \ ,\nn\\
\hat{\cal L}_4&=&\sqrt{-g}\left[-\half(\partial\hat\pi)^2\left([\hat\Pi]^2-[\hat\Pi^2]+{R\over 24}(\partial\hat\pi)^2+{R\over 2}\hat\pi[\hat\Pi]+{R^2\over 8}\hat\pi^2\right)+{R^3\over 288}\hat\pi^4\right] \ , \nn \\
\hat{\cal L}_5&=& \sqrt{-g}\left[-\half\left((\partial\hat\pi)^2+{R\over 60} \hat\pi^2\right)\left([\hat\Pi]^3-3[\hat\Pi][\hat\Pi^2]+2[\hat\Pi^3]\right)\right. \nn \\ 
&&\left.-{R\over 5} \hat\pi(\partial \hat\pi)^2\left([\hat\Pi]^2-[\hat\Pi^2]+{3R\over 16}[\hat\Pi] \hat\pi+{5R^2\over 144} \hat\pi^2\right)+{R^4\over 4320} \hat\pi^5\right] \ .
\label{singlesetGalileons}
\end{eqnarray}
Focusing on $\hat{\cal L}_2$, we note that the non-linear symmetries fix the sign of the mass term relative to that of the kinetic term.  Therefore, in de Sitter space, where $R$ is positive, the scalar is either a tachyon or a ghost, depending on the overall sign of $\hat{\cal L}_2$.  In $AdS$ on the other hand, where $R<0$, the scalar can be stable and ghost free if the sign of $\hat{\cal L}_2$ is chosen to be positive\footnote{A scalar in $AdS$ can tolerate a slightly negative mass without instability.  Any mass squared larger than the Breitenhloer Friedman bound $m^2\geq -{9\over 4L^2}={3\over 16}R$ is stable~\cite{Breitenlohner:1982bm}.  However, we cannot make use of this in any way, since the $AdS$ scalar is ghostlike whenever its mass squared is negative.}.

The presence of a tachyon suggests spontaneous symmetry breaking, as there may be higher order terms in the potential which stabilize it.  In this section, we explore the possibility of using the tachyon of the de Sitter Galileons to induce spontaneous symmetry breaking.  More specifically, consider imposing a $Z_2$ symmetry $\pi\rightarrow -\pi$, which forbids the odd terms $\hat{\cal L}_3$ and $\hat{\cal L}_5$.\footnote{This is interesting in its own right. Imposing this symmetry on the original Galileons gives an interacting scalar field theory which in suitable regimes has only one possible interaction term $\hat{\cal L}_4$, which furthermore is not renormalized.  This is the co-dimension one version of introducing an internal $so(N)$ symmetry in a theory with a multiplet of $N$ Galileons, which also yields a single possible interaction term~\cite{Hinterbichler:2010xn}.}  In the $dS$ case and $AdS$ case respectively, a symmetry breaking potential can be achieved by choosing
\bea 
\hat{\cal L}_2-a \hat{\cal L}_4,\ \ \ dS \ ,\\ 
 -\hat{\cal L}_2+a \hat{\cal L}_4.\ \ \ AdS \ ,
 \eea
with coupling constant $a>0$. In both cases, the potential is
\be 
V(\pi)={ |R|\over 288}\(-48\pi^2+aR^2\pi^4\) \ .
\ee
This has a $Z_2$ preserving vacuum at $\pi=0$ and $Z_2$ breaking vacua at $\pi=\pm\sqrt {24\over a}{1\over |R|}$.

\begin{figure} %  figure placement: here, top, bottom, or page
   \centering
   \includegraphics[width=4.0in]{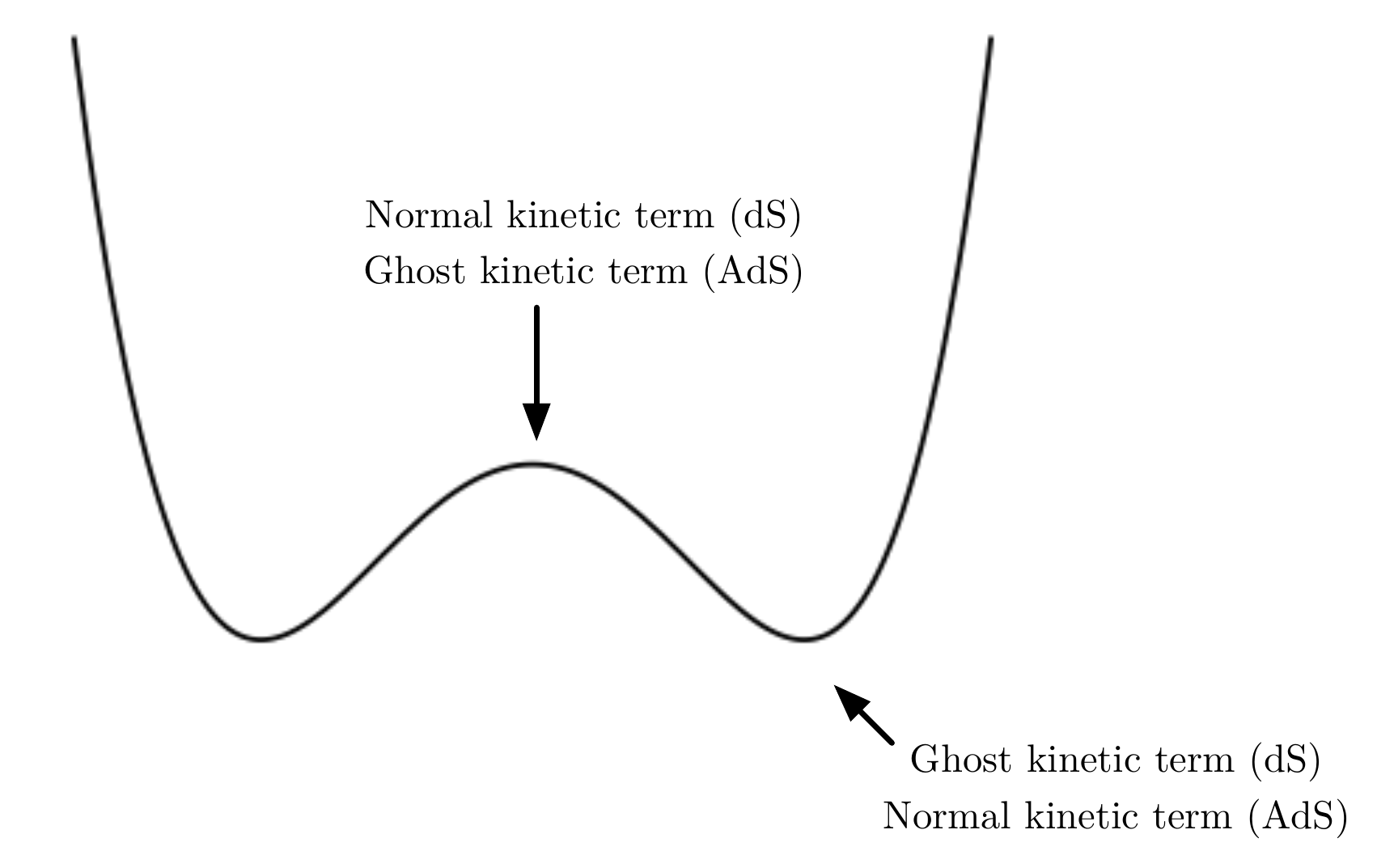}
   \caption{$Z_2$ symmetry breaking for the $dS/AdS$ Galileons.}
   \label{potential}
\end{figure}

None of these vacua alter any of the Galilean symmetries of these models.  Thus, expanding around one of the minima (the positive one, say), we obtain a Lagrangian which is also a combination of the terms~(\ref{singlesetGalileons}), with coefficients depending only on the original coefficient $a$,
\bea 
-2 \hat{\cal L}_2-\sqrt{6a} \hat{\cal L}_3-a \hat{\cal L}_4,\ \ \ dS \ ,\\ 
2 \hat{\cal L}_2-\sqrt{6a} \hat{\cal L}_3+a \hat{\cal L}_4.\ \ \ AdS \ .
\eea
In the $dS$ case, the field has a normal sign kinetic term around the tachyonic $\pi=0$ solution, and a ghostly kinetic term around the symmetry breaking vacuum.  In the $AdS$ case, the field is a ghost around the tachyonic $\pi=0$ solution, and is ghost-free around the symmetry breaking vacuum.  In this case we see a version of ghost condensation along with the usual tachyon condensation.  See figure \ref{potential}.

\section{Conclusions}
The DGP model has led to a fertile area of research, both in five dimensional braneworld models and in the $4$d effective field theories to which they lead. The Galileon theory exhibits a fascinating structure, terminating after a finite number of terms and obeying a nontrivial symmetry group arising from combinations of the higher dimensional symmetries. Perhaps most interestingly, the Galileon theories admit a non-renormalization theorem, which some authors have suggested makes them well-suited for applications to inflation and the late-time acceleration of the universe.

In this paper we have shown that the Galileon theory is a special case of a class of effective field theories that may be identified by embedding a brane solution with a general set of symmetries in a bulk of a similarly general structure. The theories obtained in this way may be interesting as examples of higher dimensional gravitating theories, or may merely provide new nontrivial examples of $4$d effective field theories.

We have derived the general conditions for the brane constructions and for obtaining the associated four-dimensional effective theories. We have then applied this construction comprehensively to all possible special cases in which both the brane and the bulk are maximally symmetric spaces	in their respective dimensionalities (with the bulk metric having only a single time direction). The results are new classes of effective field theories, sharing the important properties of the Galileons, while exhibiting distinctive new features, such as the existence of potentials, with masses fixed by symmetries. 
These potentials open up the possibility of new natural implementations of accelerating cosmological solutions in theories naturally having a de Sitter solution.

Furthermore, in some cases the potentials allow both spontaneous symmetry breaking and ghost condensation at the same time. This may allow for other new consequences of these theories, including the possibility of novel topological defects in these theories. We are currently studying these implications~\cite{GHT-newpaper}.

\bigskip
\goodbreak
\centerline{\bf Acknowledgements}
\noindent
\\
The authors are grateful to Sergei Dubovsky and Justin Khoury for discussions and comments. This work is supported in part by NASA ATP grant NNX08AH27G, NSF grant PHY-0930521, and by Department of Energy grant DE-FG05-95ER40893-A020. MT is also supported by the Fay R. and Eugene L. Langberg chair.

\appendix

\section{\label{appendix1}Some useful expressions}

Here we collect some expressions useful in the calculation leading to (\ref{generalterms}).

First some transformations in which we set $\bar g_{\mu\nu}= \tilde g_{\mu\nu}+\partial_\mu\pi\partial_\nu\pi$.  Define
\be 
\gamma={1\over \sqrt{1+\tilde g^{\mu\nu}\partial_\mu\pi\partial_\nu\pi}},\ \ \ \tilde\Pi _{\mu\nu}=\tilde \nabla_\mu\tilde \nabla_\nu\pi \ .
\ee
Brackets with tildes denote a trace with respect to $\tilde g^{\mu\nu}$, e.g. $[\tilde\Pi ]=\tilde g^{\mu\nu}\tilde\nabla _\mu\tilde\nabla _\nu\pi$, $[\tilde\Pi ^2]=\tilde g^{\alpha\mu}\tilde g^{\beta\nu}\tilde\nabla _\mu\tilde\nabla _\nu\pi\tilde\nabla _\alpha\tilde\nabla _\beta\pi$, etc. and $\left[\tilde\pi^2\right]=\tilde g^{\mu\nu}\tilde\nabla _\mu\pi\tilde\nabla _\nu\pi$, $\left[\tilde\pi^3\right]=\tilde g^{\alpha\mu}\tilde g^{\beta\nu}\tilde\nabla _\alpha\pi \tilde\nabla _\mu\tilde\nabla _\nu\pi\tilde\nabla _\beta\pi$, etc.

We have,
\be 
\bar \Gamma^\lambda_{\mu\nu}=\Gamma^\lambda_{\mu\nu}+\gamma^2\tilde\Pi _{\mu\nu}\tilde \nabla^\lambda \pi \ ,
\ee

\bea 
\bar R^\alpha_{\ \beta\mu\nu}&=&\tilde R^\alpha_{\ \beta\mu\nu}-\gamma^2\tilde R_{\gamma\beta\mu\nu}\tilde\nabla ^\gamma\pi\tilde\nabla ^\alpha\pi+2\gamma^2\left(\tilde\Pi _{[\mu}^{\ \alpha}\tilde\Pi _{\nu]\beta}-\gamma^2\tilde\Pi _{\gamma [\mu}\tilde\Pi _{\nu]\beta}\tilde\nabla ^\alpha\pi\tilde\nabla ^\gamma\pi\right) \ , \\
\bar R_{\mu\nu}&=&\tilde R_{\mu\nu}-\gamma^2\tilde R_{\alpha\mu \beta\nu}\tilde\nabla ^\alpha\pi\tilde\nabla ^\beta\pi \nn \\
&&+\gamma^2\left[\left([\tilde\Pi]-\gamma^2[\tilde\pi ^3]\right)\tilde\Pi _{\mu\nu}-\tilde\Pi ^2_{\mu\nu}+\gamma^2\tilde\Pi _{\mu\alpha}\tilde\Pi _{\nu\beta}\tilde\nabla ^\alpha\pi\tilde\nabla ^\beta\pi\right] \ , \\
\bar R&=&\tilde R-2\gamma^2 \tilde R_{\mu\nu}\tilde\nabla ^\mu\pi\tilde\nabla ^\nu\pi+\gamma^2\left([\tilde\Pi ]^2-[\tilde\Pi ^2]\right)+2\gamma^4\left([\tilde\pi ^4]-[\tilde\pi ^3][\tilde\Pi ]\right) \ .
\eea

For performing the conformal transformation, $\tilde g_{\mu\nu}=f^2 g_{\mu\nu}$, we use
\be 
\tilde\Gamma^\rho_{\mu\nu}=\Gamma^\rho_{\mu\nu}+ f^{-1}\left(\delta^\rho_\mu\partial_\nu f+\delta^\rho_\nu\partial_\mu f-g_{\mu\nu}g^{\rho\sigma}\partial_\sigma f\right) \ ,\ee
\bea \tilde R^\rho_{\ \sigma\mu\nu}&=&R^\rho_{\ \sigma\mu\nu}+2\left(-{f''\over f}+2{f'^2\over f^2}\right)\delta^\rho_{[\mu}\nabla_{\nu]}\pi\nabla_\sigma\pi-2{f'\over f}\delta^\rho_{[\mu}\nabla_{\nu]}\nabla_\sigma\pi \nn \\
&&+2\left({f''\over f}-2{f'^2\over f^2}\right)g_{\sigma[\mu}\nabla_{\nu]}\pi\nabla^\rho\pi+2{f'\over f}g_{\sigma[\mu}\nabla_{\nu]}\nabla^\rho\pi+2{f'^2\over f^2}g_{\sigma[\mu}\delta^\rho_{\nu]}(\nabla\pi)^2 \ ,\nn\\
\tilde R_{\mu\nu}&=& R_{\mu\nu}+2\left(2{f'^2\over f^2}-{f''\over f}\right)\nabla_\mu\pi\nabla_\nu\pi-2{f'\over f}\Pi_{\mu\nu}-g_{\mu\nu}\left({f'\over f}[\Pi]+\left({f''\over f}+{f'^2\over f^2}\right)[\pi^2]\right) \ ,\nn\\
\tilde R&=&{1\over f^2}R-{6\over f^3}\left(f''[\pi^2]+f'[\Pi]\right) \ .
\eea
The transformation of the matrix of derivatives is
\bea
\tilde\Pi_{\mu\nu}&=&\Pi_{\mu\nu}-2{f'\over f}\nabla_\mu\pi\nabla_\nu\pi+g_{\mu\nu}{f'\over f}[\pi^2] \ ,
\eea
and, finally, some useful relations for the contractions are
\bea 
\tilde{[\Pi]}&=&{1\over f^2}[\Pi]+2{f'\over f^3}[\pi^2]  \ ,\\
 \tilde{[\Pi^2]}&=&{1\over f^4}[\Pi^2]+2{f'\over f ^5}\left([\Pi][\pi^2]-2[\pi^3]\right)+4{f'^2\over f^6}[\pi^2]^2 \ ,\\
\tilde{[\pi^2]}&=&{1\over f^2}[\pi^2] \ ,\\
\tilde{[\pi^3]}&=&{1\over f^4}[\pi^3]-{f'\over f^5}[\pi^2]^2 \ ,\\
\tilde{[\pi^4]}&=&{1\over f^6}[\pi^4]-2{f'\over f^7}[\pi^3][\pi^2]+{f'^2\over f^8}[\pi^2]^3 \ .
\eea

\bibliographystyle{utphys}
\bibliography{generalgalileon-10}

\end{document}